\documentclass[journal]{IEEEtran}
\usepackage{amsmath,amsthm}
\usepackage{epsfig}
\usepackage{graphicx}
\usepackage{float}
\usepackage{stmaryrd}
\ifCLASSOPTIONcompsoc
  \usepackage[nocompress]{cite}
\else
  \usepackage{cite}
\fi
\usepackage[hyphens]{url}
\usepackage{color}
\usepackage[colorlinks=true,citecolor=black,urlcolor=black,linkcolor=black,hyperindex,breaklinks]{hyperref}
\usepackage[table,hyperref,x11names]{xcolor}
\usepackage{comment}
\usepackage{array}
\usepackage{hhline}
\usepackage[ruled,vlined,linesnumbered]{algorithm2e}
\usepackage{diagbox}
\usepackage{amssymb}
\usepackage{mathrsfs}
\usepackage{multirow}
\usepackage{threeparttable}
\usepackage{makecell}
\usepackage[font=small,skip=0pt]{caption}
\newcommand{\subparagraph}{}
\usepackage[hang]{footmisc}
\usepackage[normalem]{ulem}
\usepackage{scalerel}
\usepackage{pifont}

\ifCLASSOPTIONcompsoc
    \usepackage[caption=false, font=normalsize, labelfont=sf, textfont=sf]{subfig}
\else
\usepackage[caption=false, font=footnotesize]{subfig}
\fi

\theoremstyle{definition}
\newtheorem{defn}{Definition}

\newtheorem{exmp}{Example}

\newcommand{\cmark}{\ding{51}}%
\newcommand{\xmark}{\ding{55}}%
\theoremstyle{remark}

\makeatletter
\def\hlinew#1{\noalign{\ifnum0=`}\fi\hrule \@height #1
\futurelet\reserved@a\@xhline}
\makeatother

\definecolor{greyf}{rgb}{0.7, 0.7, 0.7}
\definecolor{greys}{rgb}{0.85, 0.85, 0.85}

\newcolumntype{a}{>{\columncolor{Gray}}c}
\newcolumntype{b}{>{\columncolor{white}}c}

\newcommand{\etalc}[2]{{#1 \textit{et al.}\cite{#2}}}

\newcommand{\PreserveBackslash}[1]{\let\temp=\\#1\let\\=\temp}
\newcolumntype{C}[1]{>{\PreserveBackslash\centering}p{#1}}
\newcolumntype{R}[1]{>{\PreserveBackslash\raggedleft}p{#1}}
\newcolumntype{L}[1]{>{\PreserveBackslash\raggedright}p{#1}}

\begin{document}
\title{Don't Watch Me: A Spatio-Temporal Trojan Attack on Deep-Reinforcement-Learning-Augment Autonomous Driving}
\author{Yinbo Yu, \IEEEmembership{Member, IEEE,}
        Jiajia Liu, \IEEEmembership{Senior Member, IEEE}
\thanks{Y. Yu and J. Liu are with National Engineering Laboratory for Integrated Aero-Space-Ground-Ocean Big Data Application Technology, the School of Cybersecurity, Northwestern Polytechnical University, Xi’an, Shaanxi, 710072, P.R.China (corresponding author: Jiajia Liu).}}

\maketitle

\begin{abstract}
Deep reinforcement learning (DRL) is one of the most popular algorithms to realize an autonomous driving (AD) system. The key success factor of DRL is that it embraces the perception capability of deep neural networks which, however, have been proved vulnerable to Trojan attacks. Trojan attacks have been widely explored in supervised learning (SL) tasks (e.g., image classification), but rarely in sequential decision-making tasks solved by DRL. Hence, in this paper, we explore Trojan attacks on DRL for AD tasks. First, we propose a spatio-temporal DRL algorithm based on the recurrent neural network and attention mechanism to prove that capturing spatio-temporal traffic features is the key factor to the effectiveness and safety of a DRL-augment AD system. We then design a spatial-temporal Trojan attack on DRL policies, where the trigger is hidden in a sequence of spatial and temporal traffic features, rather than a single instant state used in existing Trojan on SL and DRL tasks. With our Trojan, the adversary acts as a surrounding normal vehicle and can trigger attacks via specific spatial-temporal driving behaviors, rather than physical or wireless access. Through extensive experiments, we show that while capturing spatio-temporal traffic features can improve the performance of DRL for different AD tasks, they suffer from Trojan attacks since our designed Trojan shows high stealthy (various spatio-temporal trigger patterns), effective (less than 3.1\% performance variance rate and more than 98.5\% attack success rate), and sustainable to existing advanced defenses.

\end{abstract}

\begin{IEEEkeywords}
Backdoor attack, deep reinforcement learning, spatio-temporal feature, autonomous driving
\end{IEEEkeywords}

\IEEEpeerreviewmaketitle

\section{Introduction}

Autonomous driving (AD) is a long-researched area and a number of novel motion planning algorithms for AD have been proposed. Nevertheless, AD remains a major challenge. Traditional motion planning algorithms concentrate on the rule- or optimization-based methods \cite{kesting2007general, panwai2005comparative}. However, these rules or models are designed manually with potentially inaccurate assumptions and thus, cannot scale to different real and complex scenarios, e.g., overtaking, roundabouts, interactions, and merging. This drawback has prompted communities to turn to learning-based approaches, which bear the promise of leveraging data to automatically learn complex driving policies. Among these approaches, deep reinforcement learning (DRL), embracing the perception capability of the deep neural network (DNN) and the decision-making capability of reinforcement learning (RL), has been widely investigated for AD tasks \cite{kiran2021deep,aradi2020survey}.

DRL has been successfully used to address many sequential decision-making issues \cite{mao2017neural,wang2019autonomous, yu2022online,kiran2021deep}. Given different requirements of driving (e.g., speed, safety, and comfort), an DRL-augment AD system takes behaviors of the autonomous vehicle (AV) and surrounding human driving vehicles (HDVs) and the knowledge about road networks as input, and outputs vehicle control commands, e.g., steering, torque, and brake. The success of DRL depends on a large amount of training data and computing resources. However, the cost of human driving data collection at a large scale can be prohibitive. This dilemma can lead to the fact that, the DRL model is still not only prone to unexpected behavior in out-of-sample scenarios \cite{mo2021safe, muhammad2020deep}, but also may be injected with malicious backdoors (also called ``neural Trojan'', or ``Trojan attack'') \cite{li2020backdoor, wang2021stop}.

In this paper, we focus on the backdoor threat to DRL-augment AD through manipulating its training stage. Many existing neural Trojans can be injected into AV's capacities of vision perception \cite{li2020backdoor,gao2020backdoor}, e.g., image classification, object recognition and tracking. A backdoored DNN model behaves normally on benign samples, but can produce malicious results once an attacker-specified trigger is presented in the input (images or other sensor data). Unlike these supervised learning, DRL is required to address sequential decision-making problems according to long-term rewards instead of supervision on immediate rewards. Backdoors on DRL are more challenging since the backdoored agent needs to disrupt the sequential decisions rather than isolated decision while maintaining good performance in absence of backdoor triggers \cite{kiourti2020trojdrl}. Hence, until now, there are still a few of related works. \etalc{Kiourti}{kiourti2020trojdrl} use image patterns as the backdoor trigger and manipulate the corresponding action and reward when the trigger is present.  \etalc{Ashcraft}{ashcraft2021poisoning} studied a DRL backdoor that uses the action trajectory presented in the image observation as the trigger. \cite{wang2021stop} is the only work that studied backdoors to DRL-based AD, in which, the authors used the combinations of vehicles' speeds and positions as the trigger and studied congestion and insurance attacks in the circuit scenario. These backdoor triggers are useful for DRL-based AD and motivate us to investigate stealthier and more practice backdoors that can be easy to be activated by real-world attackers.

The DRL agent learns its optimal policy through interaction with the environment. Most DRL methods assumes that the state of the environment is fully observable for the agent. However, due to occlusions and noisy sensors of AV, the DRL agent can only glimpse a part of the system state. A partially observable Markov decision process (POMDP) \cite{kaelbling1998planning} can better capture the dynamics of the AD environment, but finding an optimal policy for POMDP is notoriously difficult \cite{hausknecht2015deep, igl2018deep}. A popular solution is to aggregate histories of observations over time by integrating a recurrent neural network (RNN) \cite{hausknecht2015deep} or a generative model \cite{igl2018deep} to find hidden states for optimal policy generation. The partial observability can make better hackers \cite{sarraute2012pomdps} and hence, motivate us that hiding a backdoor trigger in unobservable states can be stealthier.

In this paper, we aim to hide backdoor triggers into spatio-temporal traffic features which are the key factor to ensure the performance of AD \cite{pan2019urban}. First, based on the gate recurrent unit (GRU) and attention mechanism, we design a spatio-temporal DNN model for DRL to capture spatio-temporal features from observations. Through our extensive evaluations, we can find that these features can significantly improve the efficiency and safety of DRL algorithms for different AD tasks. Then, we study the security of these features by designing a novel backdoor trigger in the form of the temporal and spatial relevance of a sequence of states, rather than in a single observation used in existing neural backdoors, and proposing a spatio-temporal-pattern stealthy Trojan attack on DRL-augment AD systems. With this backdoor, the attacker can drive a vehicle in the AV's visual range, and control the spatial and temporal dependencies between the AV and her vehicle via specific driving behaviors in a short period (\textit{i.e.,} the spatio-temporal-pattern trigger) to activate the backdoor. After the trigger is present, the AV will be controlled by backdoor neurons which uses malicious rewards to generate actions. To ensure a stealthy attack, we keep the value of malicious rewards similar to the one of original genuine rewards. Compared to existing works \cite{kiourti2020trojdrl, ashcraft2021poisoning,wang2021stop}, hiding backdoor triggers into spatio-temporal dependencies further ensures its stealthy, and using temporal driving behaviors ensures the practicality of attacks. To the best of our knowledge, this is the first work that explores backdoor attacks to DRL both in the space and time domain. To evaluate our backdoor attacks, we compare our DRL algorithms with state-of-the-art (SOTA) DRL algorithms proposed for AD tasks. We evaluate their performance in several both clean and poisoned (with different triggers) driving scenarios. Our experimental results demonstrate that the performance variance rate (PVR) of backdoored models is not more than 3.1\%, and the attack success rate (ASR) is more than 98.5\%, respectively. Besides, our experiments demonstrate that SOTA defense mechanisms for DNN Trojans fail to detect our backdoor. We summarize our contributions as follows:

\begin{itemize}
    \item A spatio-temporal DRL algorithm using RNN and the attention mechanism for AD tasks;
    \item A stealthy Trojan attack on DRL algorithms using spatial and temporal features of vehicle driving;
    \item Extensive experiments with three different driving scenarios to evaluate the performance of clean and backdoored DRL models.
\end{itemize}

The remainder of the paper is organized as follows. We discuss the related work in $\S$II. $\S$III provides the detailed design of our DRL algorithm for AD tasks. We then present our proposed backdoor attacks in $\S$IV. $\S$V describes the experimental design and the performance evaluation results, and provides analytical insights. Lastly, we conclude the paper and outline our future work in $\S$VI.

\section{Related work}
\label{sec:relatedwork}

\subsection{DRL for Autonomous Driving}

An AD system is mainly composed of two parts: scene understanding and decision making \& planning \cite{kiran2021deep}. The former perceives physical driving dynamics by different sensors to provide sensor agnostic representations of the environment. The latter performs actions of steering, acceleration and braking according to routing and motion planning policies. Given the representations and routing planning, DRL is one of the most popular approaches to generate motion planning policies \cite{kiran2021deep, aradi2020survey}. It is important to design appropriate state and action spaces and rewards for DRL. While many existing methods directly use raw sensor data, (e.g., camera images, LiDAR, radar) which provides the benefit of finer contextual information, condensed abstracted data has been validated to reduce the complexity of the state space. Generally, the action space should be continuous for AD systems, including steering angle, throttle, and brake, which, however, is complex for DRL. Discretizing them uniformly by dividing the range of continuous actions into equalized bins is a feasible solution to simplify the action space \cite{kiran2021deep}.

Considering complex spatio-temporal correlations of urban traffic \cite{pan2019urban}, researchers involve RNN and attention mechanisms to design DNN models to implement AD systems.
\etalc{Mo}{mo2021safe} combined deep recurrent Q-learning (DRQN) \cite{hausknecht2015deep} and Monte Carlo tree search to implement a safe DRL for AD. For continuous controlling AV, \etalc{Chen}{chen2019attention} introduced a spatial attention network to process features extracted by CNN and output to a LSTM layer, and add temporal attention over the output of the LSTM layer. \etalc{Seong}{seong2021learning} also used a similar network architecture (spatial attention $\rightarrow$ LSTM $\rightarrow$ temporal attention), but its input is condensed abstracted data, rather than camera images. Since a general self-attention mechanisms cannot guarantee vehicle-to-vehicle attention, \etalc{Leurent}{leurent2019social} proposed a multi-head social attention mechanism into deep Q-learning (DQN), which decomposes the state space into the individual state of each observable vehicle and sends a single query generated from the AV state to the attention network. This mechanism enables the DRL agent to capture more vehicle-to-ego dependencies. \etalc{Ma}{ma2021reinforcement} focused on the effect of surrounding drivers' latent state and spatio-temporal relationships on the AD system. They provide a LSTM layer for each observable vehicle, embed these LSTM output by a graphsage convolution, and infer AV control and surrounding drivers' latent states which are merged into the next MDP sate. In this paper, we also use social attention and RNN to design a DNN to capture spatio-temporal correlations for decision making in DRL. But different from existing works, we include state, last action, and reward as input to extract more hidden temporal features by a GRU layer and feed these temporal features into the attention network to capture spatio-temporal vehicle-to-ego dependencies.

\begin{table*}[!t]
    \centering
    \caption{Comparison of existing backdoor attacks on DRL. I-T: Instant trigger. S-T: Spatial trigger. T-T: Temporal trigger. ST-T: Spatio-temporal trigger}
    \label{tab:brcomp}
    \begin{tabular}{c|ccccccc}
    \hline
        Reference & I-T & S-T & T-T  & ST-T & Action generation& Attack duration controlling & Training\\\hline
       \etalc{Kiourti}{kiourti2020trojdrl} & \cmark & \xmark& \xmark& \xmark& Fixed or Random & Instant & Single environment\\\hline

        \etalc{Yang}{yang2019design} & \cmark & \xmark & \xmark & \xmark & Reverse reward & Persistent& Two opposed environments\\\hline
        \etalc{Ashcraft}{ashcraft2021poisoning} & \xmark & \xmark & \cmark (on actions) & \xmark & Reverse reward & Instant &Two opposed environments \\\hline

        \etalc{Wang}{wang2021backdoorl} & \xmark & \xmark & \cmark (on actions) &\xmark &Reverse reward & Persistent & Two opposed environments \\\hline
        \etalc{Wang}{wang2021stop} & \cmark &\cmark & \xmark & \xmark & Specific reward & Persistent & Single environment\\\hline
        \etalc{Yu}{yu2022temporal} & \xmark &\xmark & \cmark (on states) & \xmark & Reverse reward & Controllable & Single environment \\\hline
        Our work & \xmark &\xmark & \xmark & \cmark & Specific reward & Controllable & Single environment\\\hline
    \end{tabular}
\end{table*}

\subsection{Trojans on DRL and DRL-Augment AD}

Backdoors on DNN have been widely studied in the domain of supervised learning \cite{li2020backdoor,gao2020backdoor}, where triggers can be an image patch, a physical accessory, or invisible noise. These backdoors can be applied to attack AV's vision perception capacities. For example, several invisible triggers (style transfer based \cite{cheng2021deep}, residual maps based \cite{doan2021lira}, warping based \cite{nguyen2020wanet}) are applied to a traffic sign recognition benchmark (GTSRB \cite{stallkamp2012man}). These backdoors will lead to wrong recognition of traffic signs, and then produce wrong control commands. Our work focuses on DRL-based AD. While these experiences can be applied to DRL, DRL aims to solve sequential decision-making problems which are quite different from classification tasks.

Currently, there are only few works that have studied backdoors on DRL. We compare these works in Table \ref{tab:brcomp}.
\etalc{Kiourti}{kiourti2020trojdrl} specify image patterns as the trigger and train a DRL model which can output fixed or random actions or reward when the trigger is present. \cite{wang2021stop} is the only work that studied backdoor attacks on DRL-based AV controllers. The authors used a specific set of combinations of positions and speeds of vehicles in the observation as the trigger. When the trigger is present, the backdoored DRL model could generate malicious decelerations to cause a physical crash or traffic congestion. We call these two backdoors ``instant backdoors'', similar to the one in supervised learning, because their triggers and actions do not involve sequential dependencies.

Since the state space of many real-world environments including AD is partial observable, introducing RNN into DRL networks to capture temporal dependencies is an efficient solution to improve the performance and practicability of DRL. But, RNN can introduce new security threats to DRL. Similar to \cite{kiourti2020trojdrl}, \etalc{Yang}{yang2019design} used image patches as the trigger, but they examined the persistent effect of backdoors on DRL equipped with LSTM by jointly training the backdoored model with a clean and a Trojan environment. \etalc{Ashcraft}{ashcraft2021poisoning} also studied backdoors on a network similar to the one in \cite{yang2019design}, but the trigger they used is a sequence of actions following a specific pattern, and their attacks are instant.  \etalc{Wang}{wang2021backdoorl} studied backdoors on DRL controlled two-player competitive games. They also used a sequence of actions as the fast-failing trigger, trained two different policies (clean and poisoned), and hard-encoded them to generate the backdoored trajectories. Then, they used imitation learning to train a DNN model to mimic the behavior of generated trajectories. Our previous work \cite{yu2022temporal} proposed a temporal-pattern backdoor attack on DRL, which uses the temporal constraints among a sequence of states (which are not controllable by the DRL agent) to encode its trigger. In this paper, we design a new backdoor attack on DRL by hiding triggers in spatio-temporal features, which are the key success factor of DRL-based AD systems, and train the backdoored model in the same environment, rather than two individual environments. Besides,  different from existing backdoors, the attack duration after the trigger is present is controllable during the training process, which balances the low low poisoning rate of training buffers and high attack success rate.

\subsection{Defense}

Most existing DNN backdoor defense mechanisms are geared towards classification networks \cite{li2020backdoor,gao2020backdoor}.
\etalc{Tran}{tran2018spectral} studied how to filter poisoned samples from the training set and demonstrated that poisoned samples tend to contain detectable traces in their feature covariance spectrum.
STRIP \cite{gao2019strip} can detect whether an input contains Trojan trigger by adding strong perturbation to the input. Neural Cleanse \cite{wang2019neural} identifies backdoored DNN models through reversing engineers for each image input, so that all inputs stamped with the pattern are classified to the same label. DeepInspect \cite{chen2019deepinspect} learns the probability distribution of potential triggers from the queried model and retrieves the footprint of backdoors. These defense mechanisms are designed for isolated inputs rather than sequential inputs. Recently, \etalc{Guo}{guo2022backdoor} proposed a backdoor detection method in competitive DRL \cite{wang2021backdoorl} by training a separate policy with a reversed reward function given by the Trojan agent. Since our backdoor triggers are hidden in a series of sequential inputs, rather than a single input, retraining models is more hard to cover the trigger pattern. More sophisticated defense methods for DRL backdoor are desired to develop.

\section{Spatio-Temporal DRL-Augment AD}
\label{sec:stad}
The task of lane change decision making and speed controlling for AV can be formulated as a Markov decision process (MDP). However, due to occlusions and noisy sensors, the AV can only glimpse a part of the surrounding state. Hence, we build AD tasks as a POMDP. Considering the spatio-temporal traffic features, we design a deep spatio-temporal reinforcement learning algorithm for autonomous driving.

\subsection{Partially-Observable Markov Decision Process}
\label{sec:pomdp}
Formally, a POMDP can be described as a 5-tuple $(\mathcal{S},\mathcal{A},\mathcal{O},\mathcal{P}, \mathcal{R})$, where $\mathcal{S},\mathcal{A},\mathcal{O}, \mathcal{R}$ are the state space, action space, observation space, and rewards. In general, at each epoch $t$, the DRL agent observes a state $s_t \in \mathcal{S}$ including the features of driving information about the AV and HDV, and then selects an action $a_t \in\mathcal{A}$ (\textit{i.e.,} control signal) to the AV.  After performing $a_t$, the AV will return a reward $r_t\in\mathcal{R}(s_t,a_t)$ to the agent. After, the agent obtains the next state $s_{t+1}$ following the transition function $\mathcal{P}(s_{t+1}|s_t,a_t)$. Due to partial observability, the agent actually can only observe the partial state, \textit{i.e.,} $s_t\in \mathcal{O}$, where $\mathcal{O}$  is the partial observation space.

\begin{figure}[t!]
\centering
\includegraphics[width=1\columnwidth]{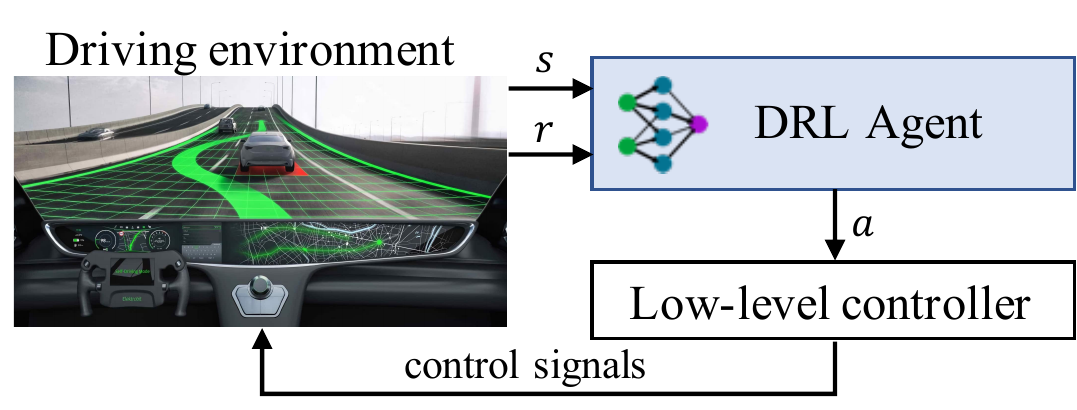}
\caption{The architecture of learning-augment autonomous driving system.}
\label{fig:arch}
\end{figure}

\textbf{State Space}: In general, the features of a vehicle driving can be represented by its continuous position, velocity, and heading. In this paper, we assume there are $I-1$ observable HDVs in the vicinity of the AV, and use a modified kinematic bicycle model \cite{kong2015kinematic} to describe the features of the AV and other observable HDVs as follows:

\begin{eqnarray}
  s_t=\left(s_{i}\right)_{i \in[0, I)}, \ \text{where}\qquad\qquad\qquad\qquad\qquad\qquad\quad \\\nonumber
  s_{i}=\left[\begin{array}{lllllll}
  p & x_{i} & y_{i} & v_{i}^{x} & v_{i}^{y} & \cos \psi_{i} & \sin \psi_{i}
  \end{array}\right]^{T}.
\end{eqnarray}

\noindent For the AV (\textit{i.e.}, $i=0$), $x_{i}$ and $y_{i}$ are the longitudinal and lateral position of $i$ in the road, respectively; $v_{i}^{x}$ and $v_{i}^{y}$ are the longitudinal and lateral speed of $i$, respectively; and $\cos\psi_{i}$ and $\sin\psi_{i}$ are the cosine and sine of heading $\psi_{i}$ of $i$, respectively. For other HDVs (\textit{i.e.}, $i>0$), these variables of position ($x_{i}$ and $y_{i}$) and velocity ($v_{i}^{x}$ and $v_{i}^{y}$) are relative to the AV, e.g., $v_i^x=v_i^x-v_0^x$. Besides, we introduce a binary variable $p$ for each vehicle that indicts whether it is observable in the vicinity of the AV. This representation uses the smallest quantity of information necessary to represent the scene and has been proved to be efficient for AD tasks \cite{bai2015intention, chen2017socially, leurent2019social}.

\textbf{Action Space}: The DRL agent aims to take suitable actions for motion planning, including lane keeping, navigation, simple racing, overtaking, or maneuvering tasks. These tasks can be completed by longitudinal (speed changes) and lateral (steering control) actions. Hence, we use five high-level control decisions, including \textit{turn left}, \textit{cruising}, \textit{turn right}, \textit{speed up}, and \textit{slow down}, to construct the action space following the designs in \cite{leurent2019social, chen2020autonomous}. A high-level action is then translated into the corresponding steering and throttle control signals to maneuver the AV by the lower-level vehicle controller.

\textbf{Reward}: The reward function is used to optimize the DRL policy and composed of three terms:
\begin{itemize}
  \item \textbf{Collision evaluation} $r_c$: if the AV is crashed, $r_c$ is set to -1, otherwise $r_c=0$;
  \item \textbf{Stable-speed evaluation} $r_s$: $r_s$ is used to reward faster speed $v_t$ within the speed limit $[v_{min}, v_{max}]$:
  \begin{equation}
    r_s= \min\{\frac{v_t-v_{min}}{v_{max}-v_{min}}, 1\}.
  \end{equation}
  The speed limit may vary in different roads, e.g., highways in China has the limit $[22\ \text{m/s}, 33\ \text{m/s}]$, and the speed range observed in the Next Generation Simulation (NGSIM) dataset\footnote{https://ops.fhwa.dot.gov/trafficanalysistools/ngsim.htm} is $[6\ \text{m/s}, 17\ \text{m/s}]$. Hence, we set the pair of $v_{max}$ and $v_{min}$ according to the road.
  \item \textbf{Time-headway evaluation} $r_h$: to further prioritize safety, we evaluate the time headway of the AV as:
  \begin{equation}
    r_h=\log{\frac{d_{h}}{t_hv_t}},
  \end{equation}
  where $d_h$ is the distance headway; and $t_h$ is a time headway threshold and is set to 1.2 $s$ as suggested in \cite{ayres2001preferred}. With $r_h$, the AV will get rewarded when $d_h$ is greater than $t_hv_t$, otherwise get penalized.
\end{itemize}

Given these metrics, we combine them as a single objective for training the DRL policy at time step $t$ as follows:

\begin{equation}
\label{equ:nrew}
  r_t = w_cr_c+w_sr_s+w_hr_h,
\end{equation}
\noindent where $w_c$, $w_s$, and $w_h$ are positive weighting scalars.

\subsection{Deep Reinforcement Learning}
Given the POMDP, we envisage the agent with the goal of maximizing the expectation of the long-term rewards. Starting at an initial state $s_i$, the agent would learn a policy $\pi(a|s)$, which maps each state $s$ to an action $a$ so that the selected action can maximize the expected rewards of all consecutive steps after $s_i$. The policy can be evaluated by its \textit{action-value} function, also referred as Q-function:
\begin{equation}
  Q(s, a) = \mathbb{E}_{\pi}\left[\sum_{k=0}^{\infty} \gamma^{k} R\left(s^{(t+k)}, \pi\left(s^{(t+k)}\right)\right) \mid s^{(t)}=s_i\right],
\end{equation}

\noindent where $\gamma\in [0,1]$ is the discount factor and specifies the importance between future rewards and the current reward, \textit{i.e.}, $\gamma=0$ represents a ``myopic'' agent only concerned with its immediate reward, while $\gamma=1$ denotes an agent striving for a long-term higher reward. Using the Bellman equation and temporal difference, we can simplify the Q-function for policy learning as:

\begin{equation}
  Q_t(s,a)=Q_{t-1}(s,a)+\alpha\left(r+\gamma\max_{a'}Q(s',a')-Q_{t-1}(s,a)\right),
\end{equation}
\noindent where $\gamma$ is the learning rate; $a'$ maximizes all possible $Q(s',a')$. So we can take the previous $Q_{t-1}(s,a)$ and add on the temporal difference times $\gamma$ to get the new $Q_t(s,a)$.

\begin{figure}[t!]
\centering
\includegraphics[width=1\columnwidth]{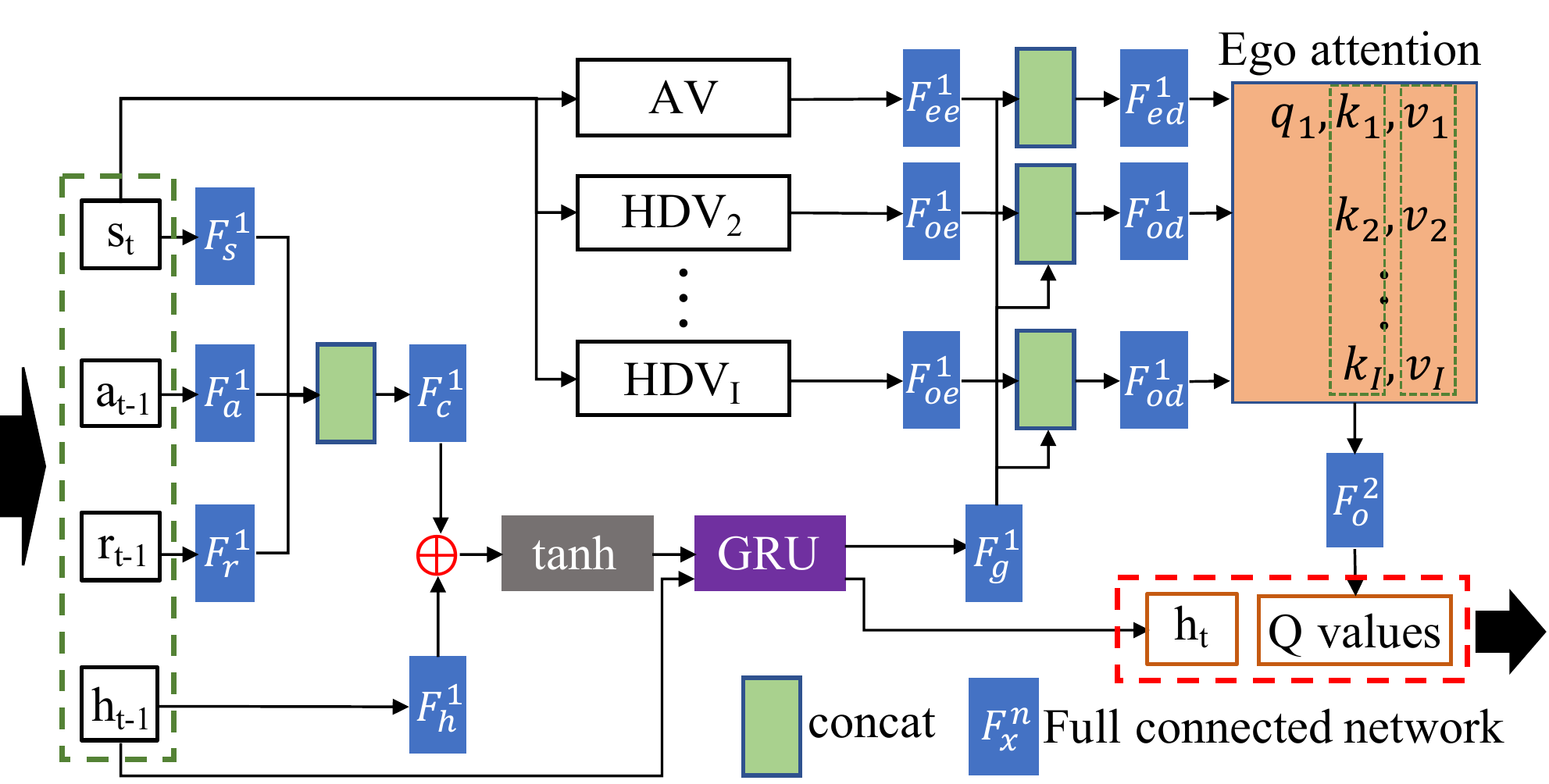}
\caption{Spatio-Temporal DNN for Autonomous Driving.}
\label{fig:net}
\end{figure}

With the Q-function, we can learn the policy $\pi$ by the DQN algorithm. But since the AD environment is partially observable, finding an optimal policy for POMDP is notoriously difficult. Integrating recurrent neural networks (RNN) to find hidden state over time is an efficient solution for addressing POMDP problems \cite{hausknecht2015deep}. Existing literature \cite{pan2019urban} has shown that traffic models and their changing patterns have high spatio-temporal correlations. Hence, in this paper, based RNN and attention mechanism, we design a spatio-temporal DNN for learning the DRL policy, as shown in Fig. \ref{fig:net}. With this network, we call our DQN algorithm \textbf{DAGQN} and introduce this network in detail as follows.

First, the historical driving patterns of different vehicles and the dynamic changes of traffic conditions ahead over time are very critical for reasonable motion decision making \cite{kiran2021deep}. We use RNN to aggregate histories of observations over time and identify these temporal features for optimal policy generation. As studied in \cite{ni2021recurrent}, inputting more running information (e.g., observation, action, and reward) into RNN can yield a policy with better performance. Hence, the inputs of our DAGQN contain current observation $s_t$, previous executed action $a_{t-1}$ and obtained reward $r_{t-1}$. We concatenate features extracted from $s_t$, $a_{t-1}$, and $r_{t-1}$ into one dimensional tensors $f_{c}$. Besides, similar to the soft attention proposed in \cite{sorokin2015deep}, we further add the feature extracted from previous hidden state $h_t$ to $f_{c}$ for capturing more historical information. Finally, an gated recurrent unit (GRU) layer is used to capture temporal information. The calculation process of this part is formulated as follows:
\vspace{-2mm}

 \begin{gather}\nonumber
 f_{c}=\langle F_s^1(s_t),F_a^1(a_{t-1}),F_r^1(r_{t-1}), \\\nonumber
 f_g, h_t =\text{GRU}(\text{tanh}\left(F_c^1(f_{c})+W_{h}*h_{t-1}\right), h_{t-1}),\\\nonumber
 f_g' =F_g^1(f_g),
 \end{gather}

\noindent where $\langle \rangle$ means concatenation; $F_x^n$ is a full connected network (FCN) with $n$-layer linear networks, each of which is an affine transformation with weights matrix $W$ and bias $b$; the bias of $F_h^1$ is set to be 0, \textit{i.e.}, $F_h^1(h_{t-1})=W_h*h_{t-1}$; besides $F_h^1$, all linear layers are followed by a relu activation function. The output temporal feature $f_g$ is input by a linear layer $F_g^1$ and a relu activation function.

Second, spatial positions of vehicles and their interaction behaviors are another type of information for motion planning. To tackle these spatial features, filtering out relevant spatial information from them and capturing vehicle-to-ego dependencies is necessary for efficient decision making. Hence, DAGQN uses a multi-head social attention mechanism to enforce the agent to pay more attention to these HDVs which affect the planned route of the AV. In the attention network, we first split $s_t$ into individual state $s_t^i$ of each AV and HDV ,and then extract the feature from them by linear layers ($F_{ee}^1$ for AV and $F_{oe}^1$ for other HDVs) and concatenate the feature with extracted temporal features $f_g'$. Then, we encode these individual features into the embedding with the size $d_x$ by linear layers $F_{ed}^1$ and $F_{od}^1$, and feed them to an ego-attention layer proposed by \cite{leurent2019social}. In the ego-attention layer, the AV emits a single query $\mathcal{Q}=[q_1]\in \mathbb{R}^{1\times d_k}$ by a linear projection $L_q\in \mathbb{R}^{d_x\times d_k}$, and all vehicles emit a key set $K=[k_1,\cdots,k_I]\in \mathbb{R}^{I\times d_k}$ and a value set $V=[v_1,\cdots,v_I]\in \mathbb{R}^{I\times d_k}$ by two linear projections $L_k\in\mathbb{R}^{d_x\times d_k}$ and $L_v\in\mathbb{R}^{d_x\times d_v}$, respectively. In each head, the query is compared to each key $k_i$ by their dot product $q_1k_i^{T}$, scaled by the inverse-square-root-dimension $1/\sqrt{d_k}$, normalized with a softmax function $\sigma$ across vehicles, and finally is used to gather the value $v_i$. The outputs from all heads are combined with a linear layer as the final Q value. The calculation process for attention computation is formulated as follows:
\vspace{-2mm}
\begin{gather}\nonumber
  f_1=F_{ed}^1(\langle F_{ee}^1(s_t^1),f_g'\rangle),\\\nonumber
  f_i=F_{od}^1(\langle F_{oe}^1(s_t^i),f_g'\rangle), i>1, \\\nonumber
  \mathcal{Q}=[L_q(f_i)], K=L_k(f_1,\cdots,f_I), V=L_v(f_1,\cdots,f_I),\\\nonumber
  \text{output}=\sigma\left(\frac{\mathcal{Q} K^{T}}{\sqrt{d_{k}}}\right)V,\\\nonumber
  Q =F_{o}^2(\text{output}).\nonumber
\end{gather}

\section{Spatio-Temporal Backdoor Attack}

\subsection{Threat model}

\textbf{Attacker’s Capacities}. Similar to existing DNN backdoors \cite{wang2021stop,kiourti2020trojdrl,ashcraft2021poisoning}, we consider the outsourcing scenario, in which the large scale model training is outsourced to a third party due to lack of computational resources or AD environments. In this scenario, attackers are allowed to poison some training data, control the training process, and manipulate rewards to poisoned states. This security threat exists in many DNN-based applications, including DRL-augment AD \cite{gao2020backdoor,gong2021defense}.

\textbf{Attacker’s Goals}. With these above capacities, backdoor attackers aim to embed hidden backdoors into DRL policies. In particular, these inserted backdoors should have effectiveness, stealthiness, and sustainability \cite{gao2020backdoor}. The effectiveness requires that the backdoored model can perform no different from a normally-trained model in absence of backdoor triggers and generate desired malicious actions when the trigger is present; the stealthiness requires that backdoor triggers should be concealed and have a small poisoning rate; the sustainability requires that the attack should still be effective under some common backdoor defenses.

\subsection{Spatio-Temporal-Pattern Trigger}

The AD system is a typically input-driven system that not all observations is controllable for DRL decisions and even may be not observable, e.g., surrounding drivers' latent behaviors \cite{ma2021reinforcement}. These states may be possible to be manipulated by malicious users or attackers. For example, the attacker drives a HDV with a specific behavior which follows the predefined backdoor trigger to the DRL-based AV controlling system in the AV. In this paper, we aim to design a backdoor attack on the DRL-based AV, whose triggers are hidden in these states both from the spatial and temporal forms, we called \textit{spatio-temporal-pattern trigger}. Such a backdoor attack is stealthy and realistic since the attacker only need to drive a HDV around the AV. Here, we define the trigger of this backdoor attack as follows:

The state space defined in Section \ref{sec:pomdp} can be divided into two parts: the states of the AV are controllable for the DRL agent; the states of neighboring vehicles are uncontrollable. For reality, we assume the attacker drives a HDV neighboring the AV, and we embed backdoor triggers both in the spatial dependencies between the attacker's HDV and the AV and in the temporal behaviors of the attacker's HDV. In our previous work \cite{yu2022temporal}, we defined a temporal trigger in the form of temporal logic formula over a sequence of time series states. While the spatio-temporal trigger can be defined in the formula, this formula lacks of intuitiveness of attack implementation. Hence, we formulate its spatial dependencies (including velocity, position, and heading) in the logic formula, but its temporal behavior in the form of a set of vehicle controlling commands. Given a state $s_e$ of the AV and $s_a$ of the attacker's vehicle, we define a formula $\varphi := f(s_e, s_a)\sim c$ to describe a constraint over $s_e$ and $s_a$, where $f$ is a binary operation (e.g., $+, -, \times, \div$), $\sim\in\{\equiv,\neq, >, \geq, <, \leq\}$, and $c\in\mathbb{R}$ is constant. The syntax of a spatio-temporal-pattern trigger is defined as follows:

\begin{defn} Given a set $S$ of states over the period length $\mathbb{N}^a$ ending at timestep $t$, a \textbf{spatio-temporal-pattern trigger} $\tau:=(\phi,\xi)$ is defined by a logic formula $\phi$ over spatial constraints $\Psi$ of these states and a set $\xi$ of controlling commands:
\begin{eqnarray}
  &\phi := \varphi_{(i,j)}|\varphi_{(i,j)} \otimes \varphi_{(k,s)}|ite(\varphi_{(i,j)}, \varphi_{(k,s)}, \varphi_{(p,q)}),\\
  &\xi := (a_{t-\mathbb{N}^a+1},a_{t-\mathbb{N}^a+2},\cdots,a_t),
\end{eqnarray}
\end{defn}
\noindent where $\varphi_{(i,j)}, \varphi_{(k,s)}, \varphi_{(p,q)} \in\Psi$, $i, k, p\in\mathbb{N}$, and $j,s,q$ are the features of vehicle-to-ego; $\otimes$ is a Boolean operator, including disjunction $\vee$ (``or'') and conjunction $\wedge$ (``and''); $ite$ denotes $\phi$-assignment, e.g., $\phi := ite(\varphi_1, \varphi_2, \varphi_3)$ means if $\varphi_1$ is true, $\phi := \varphi_2$; otherwise $\phi := \varphi_3$. $a$ is an action in the action space performed in the attacker's vehicle. With the formula, attackers can define a backdoor trigger over a set of states in a period $\mathbb{N}$ and drive following the specified behavior.

\begin{exmp}
\label{exm:trigger1}
For instance, given a trigger ($\mathbb{N}^a=4$):
$\phi := \varphi_{(t-3, p)}\wedge\varphi_{(t-3,l)}\wedge\varphi_{(t-2, v)}\wedge\varphi_{(t-2, l)}\wedge\varphi_{(t-1, v)}\wedge\varphi_{(t-1, l)}\wedge\varphi_{(t, v)}\wedge\varphi_{(t, l)}$, $\xi := (a_{t-3},a_{t-2},a_{t-1})$, where $\varphi_{(t-3, p)}:= p^a_{t-3}-p^e_{t-3}>30\wedge p^a_{t-3}-p^e_{t-3}<60$, $\varphi_{(t-3, l)}:= l^a_{t-3}-l^e_{t-3}=1$, $\varphi_{(t-2, v)}:= v^a_{t-2}>24.5\wedge v^a_{t-2}<25.5$, $\varphi_{(t-2, l)}:= l^a_{t-2}-l^e_{t-2}=0$, $\varphi_{(t-2, v)}:= v^a_{t-1}>24$, $\varphi_{(t-1, l)}:= l^a_{t-1}-l^e_{t-1}=0$, $\varphi_{(t, v)}:= v^a_{t}<20.5\wedge v^a_{t}>19.5$, $\varphi_{(t, l)}:= l^a_{t}-l^e_{t}=1$, $a_{t-3}$ includes turning to left lane and accelerating to 25 m/s, $a_{t-2}$ is to cruise, and $a_{t-1}$ includes turning to right lane and decelerating to 20 m/s. $p$ means the position on the lane, $l$ is the lane index, and $v$ is the velocity of a vehicle. $l^a-l^e=1$ means the attacker's vehicle is on the right lane of AV and $ l^a-l^e=0$ means they are on the same lane.
\end{exmp}

In this example, the attack's HDV is on the front right lane ($\varphi_{(t-3, l)}$) of AV with the distance defined by $\varphi_{(t-3, p)}$. Then, the attack's HDV turns left and accelerate to 25 m/s. After getting in the same lane of the AV, it cruises for a timestep. Finally, it turns right and decelerates to 20 m/s. After observing such a specific driving behavior, the AV will be controlled by the backdoored neurons.

\subsection{Backdoored Model Training}

Our attack aims to cause the AV to crash into the attacker's HDV or other HDVs with the goal of making traffic congestion or insurance claims. This attack can be accomplished by inducing the AV to perform malicious actions, e.g., accelerate if it is close to its leader. Wang \textit{et. al.} \cite{wang2021stop} proposed an insurance attack for driving on a circular track, in which the AV crash can be implemented by tricking the AV to keep acceleration since the objective of the attack is to crash the AV into its front HDV. Our goal is more general that we aim to crash into any possible HDVs within a shorter time. We implement our backdoor attack using the guidance of designed malicious rewards. In this section, we present how to train the backdoored DRL model for AD tasks.

\begin{algorithm}[!t]
\caption{Spatio-Temporal-Pattern Backdoored DAGQN.}\label{alg:tpb}
\linespread{0.95}\selectfont
\small
Initialize replay memory $D$\ to capacity $M$, temporal-pattern trigger $\tau(\phi,\xi)$, backdoor duration $L$, poisoning rate $\lambda$\;
Initialize Q-Network $Q$, Target-Network $\hat{Q}=Q$ \;
\For{$episode =1 $ \text{to}$\ max\_training\_steps$}{
With probability $\lambda$ enable the malicious vehicle injection\;
$poisonDur =0$\;
\For{$t=1$ \text{to} $\mathbb{N}$}{
  With probability $\epsilon$ select a random action $a_t$, otherwise select $a_t = \max_a Q(s_t,a)$\;
  Execute action $a_t$ and observe reward $r_t$, new state $s_{t+1}$, and terminal signal $d_t$\;
  \If{$\tau$ is active and $t$ is the ending step of $\mathbb{N}^a$}{
    $poisonDur = L$\;
  }
  \If{$poisonDur>0$}{
     $r_t = r^a_t$\;
     $poisonDur = poisonDur -1$\;
  }
  Store $(s_t, a_t, r_t, d_t, s_{t+1})$ into $D$\;
  Sample sequential batches $(s_i, a_i, r_i, d_i, s_{i+1})$ from $D$\;
  Calculate target $y_i= r_i+\gamma\max_{a'}Q(s_{i+1},a')$\;
  Do a gradient descent step with loss $\|y_i-Q(s_i,a_i)\|^2$\;
  Every $C$ steps reset $\hat{Q}=Q$\;
}
}
\end{algorithm}
\setlength{\textfloatsep}{0pt}%

First, the attacker specifies a spatio-temporal-pattern trigger $\tau:=(\phi, \xi)$ with a trigger period $\mathbb{N}^a$. Note that although with GRU, the DRL model can have long memory effects, GRU is not that ideal and require more complex neurons and longer training time for memorizing longer historical features \cite{zhao2020rnn}. Solutions, like \cite{igl2018deep, zhao2020rnn}, can address this limitation and thereby, make our backdoor attacks have longer effects, but are out of our scope. Hence, we keep only $\mathbb{N}^a$ no more than 10 time steps. To generate malicious actions after the trigger is present, we design a backdoor reward $r^a$ to replace the original reward. It is composed of two terms: (1) \textbf{Collision evaluation} $r_c^a$: if the AV is crashed, $r_c^a$ is set to 1, otherwise $r_c^a=0$; (2) \textbf{Headway evaluation} is used to reward the smaller distance between the AV and its front:
\begin{equation}
\label{equ:harew}
  r_h^a = \frac{d^{max}-\text{clip}(d_h,0,d^{max})}{d^{max}},
\end{equation}
\noindent where $d^{max}=60$m. With two positive weighting scalars ($w_c^a$ and $w_h^a$), the backdoor reward is as follows:

\begin{equation}
\label{equ:arew}
    r^a=w_c^ar_c^a+w_h^ar_h^a.
\end{equation}
\noindent To achieve a \textit{stealthy} attack, we keep the malicious reward function having values similar to the genuine one by setting suitable $w_c^a$ and $w_h^a$, but it can guide the agent to generate malicious actions.

Given the backdoor reward $r_c^a$, the DAGQN model can be trained to contain backdoored neurons. However, keeping $r_c^a$ persistent will make the replay memory used in the training process poisoned excessively, so that a low PVR of the backdoored model cannot be guaranteed. Besides, persistent $r_c^a$ may make the backdoor detection relatively easy. Hence, to ensure the stealthy (\textit{i.e.,} low poison rate) of our backdoor attacks, we introduce an attack duration $L$ to control the period of generating malicious actions after the trigger is present.

With the designed $\tau(\phi,\xi)$, $L$, and $r_c^a$, we train the backdoored DAGQN model for AD tasks following the algorithm \ref{alg:tpb}. To keep the balance between PVR and ASR of the backdoored model, we set a poisoning probability $\lambda<1$ for each training iteration. During each training iteration, the DRL agent is trained in the off policy manner, that it selects an action at each timestep using the $\epsilon$-greedy method (Line 7), stores interactions (Line 8 and 14), and updates the network with a MSE loss $\|y_i-Q(s_i,a_i)\|^2$ (Line 14-18). For an iteration that contains a malicious HDV, once the HDV meets with these spatial constraints of the first timestep of $\phi$, it will drive following the $\xi$. At the end of $\mathbb{N}^a$, we change the reward function from the one defined in Equ. (\ref{equ:nrew}) to $r^a$. This setting will lead the network to being updated in the objective of crashing the AV, thereby is undesired. Considering our backdoor trigger is spatio-temporal, to update the network, randomly sampling batches from the memory $D$ will break original sequential features. Hence, we sample a sequential batch from the memory $D$ to help the DRL model learn more sequential features and also ensure the ASR of our backdoor attacks.

\section{Evaluation}
\label{sec:experiment}
In this section, we first compare the performance of our proposed algorithm to SOTA in terms of driving safety (\textit{i.e.,} collision rate) and efficiency (\textit{i.e.,} velocity). Then, we evaluate the PVR and ASR of our backdoor attacks to these algorithms in different driving scenarios.

\subsection{Experiment Settings}

\begin{figure}[t!]
\centering
\begin{minipage}[c]{1\columnwidth}
    \centering
    \subfloat[Highway]{\includegraphics[width=1\columnwidth]{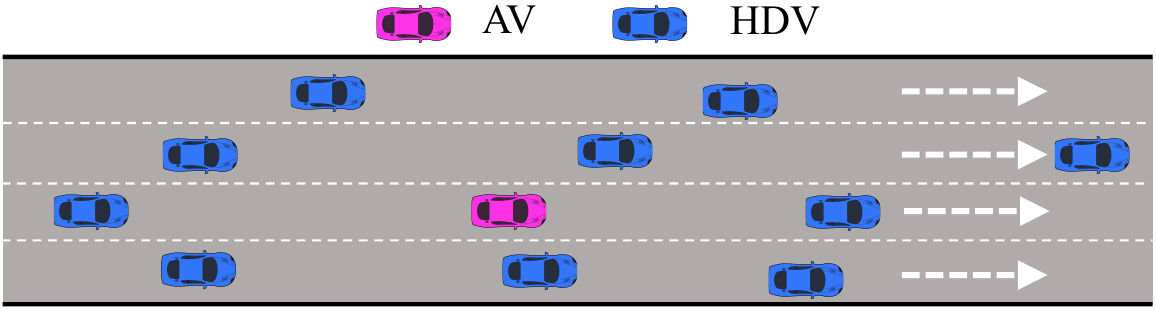}}
  \end{minipage}

  \begin{minipage}[c]{1\columnwidth}
    \centering
    \subfloat[Two-way]{\includegraphics[width=1\columnwidth]{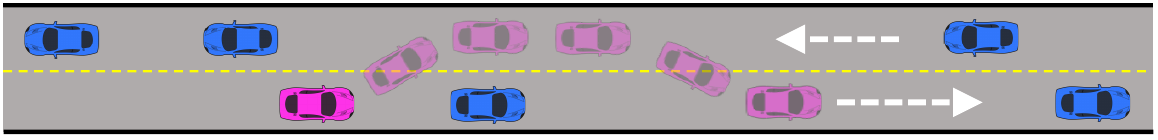}}
  \end{minipage}

  \begin{minipage}[c]{1\columnwidth}
    \centering
    \subfloat[Roundabout]{\includegraphics[width=0.8\columnwidth]{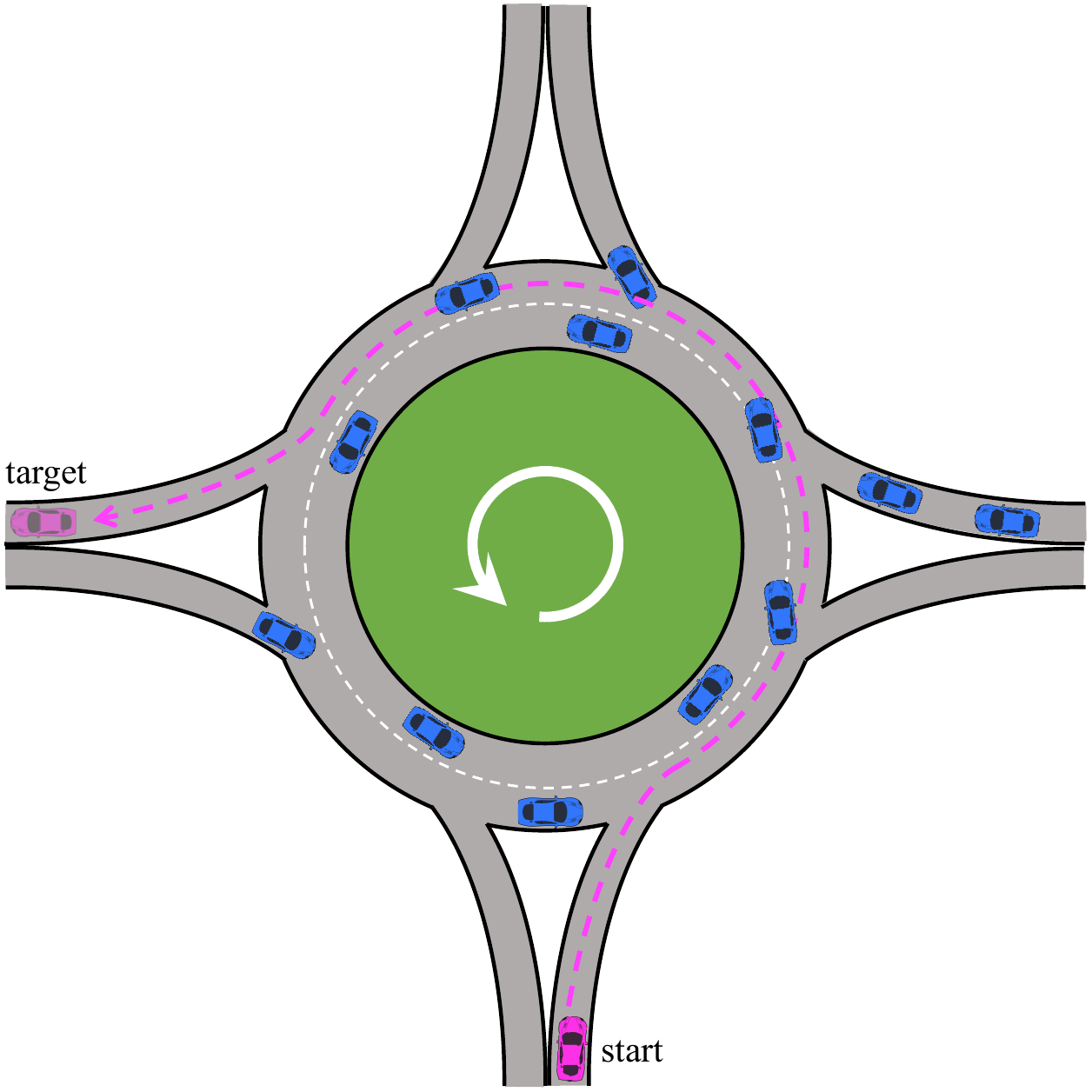}}
  \end{minipage}
  \caption{Scenarios.}
    \label{fig:scen}
\end{figure}
To evaluate driving algorithms and our backdoor attacks, we use three typical autonomous driving scenarios: highway, two-lane overtaking, and roundabout (see Fig. \ref{fig:scen}). In the highway scenario shown in Fig. \ref{fig:scen} (a), we set four lanes in one direction, the road length to be 10000m, the allowed speed range to be 20 to 30m/s. The initial speed $v_0$ of all vehicles is randomly chosen between 21 to 27 m/s. We randomly generate 40 HDVs and 1 AV along the road by choosing random lane ID ($l^{ID}$) from $[0,3]$ and longitudinal position gap between their initial spawn positions from a uniform distribution $[0.9*(12+v_0)e^{-n^{lane}/8},1.1*(12+v_0)e^{-n^{lane}/8}]$, where $n^{lane}$ is the number of lanes. In the two-way scenario shown in Fig. \ref{fig:scen} (b), we set the road length to be 1500m and the speed limit to be 16m/s. In the following ($l^{ID}=0$) and incoming $l^{ID}=1$ lane, we both generate 10 random HDVs, but their initial speed range is $[6,10]$m/s and $[4,12]$m/s, respectively, and they have a different spacing parameter $sp^0=5$ and $sp^1=7$ for spawn position generation from $[0.9*sp*(12+v_0)e^{-n^{lane}/8},1.1*sp*(12+v_0)e^{-n^{lane}/8}]$. The AV in the following lane needs to overtake the front HDVs with consideration of the driving conditions of HDVs in the incoming lane. In the roundabout scenario shown in Fig. \ref{fig:scen} (c), we have two circular lanes with radius 60m ($l^{ID}=0$) and 64m ($l^{ID}=1$), respectively. In the four directions of east, west, north and south, there is one entrance and exit lane, respectively, connecting to the lane $l^{ID}=1$. In each entrance lane, we randomly generate 3 HDVs with the space gap distribution $[10,60]$m. In the circular lanes, we randomly generate 6 HDVs with a spacing parameters $sp=2$. The initial speed range of HDVs in entrance and circular lanes is $[9,15]$ and $[7,13]$m/s, respectively.

We modify the gym-based highway-env simulator \cite{highwayenv} to simulate the environment for these above driving scenarios. We use the default parameters of the IDM and MOBIL used in the highway-env simulator to implement longitudinal and lateral controlling of HDVs. The control sampling frequency both of HDVs is 5 Hz, \textit{i.e.}, a HDV takes an action every 0.2 seconds. We set that there are 6 front HDV observable for the AV, \textit{i.e.,} $I=7$. For the DRL model, we have the following network size settings: $F_s^1=7I\rightarrow64$, $F_a^1=F_r^1=1\rightarrow8$, $F_c^1=80\rightarrow128$, $F_h^1=F_g^1=128\rightarrow128$, $F_{ee}^1=F_{oe}^1=7\rightarrow64$, $F_{ed}^1=F_{od}^1=192\rightarrow128$, and $F_o^2=128\rightarrow128\rightarrow5$. To train the model, we set the capacity of replay memory $M=60000$, the size of batch 4 for sequential batches, the sequential length of each batch less than 30, discount factor $\gamma=0.9$, and steps of updating target network $C=50$, and the learning rate of $\alpha=0.001$. The coefficient $w_c$, $w_s$, $w_h$, $w_c^a$, and $w_h^a$ for the reward function are set as 1.2, 0.4, 0.4, 2.5, and 0.4, respectively. Besides, we set $\epsilon$ to be decreased from 0.9 by 0.002 during network training.

To train the DRL model with backdoors, besides using the trigger defined in Example \ref{exm:trigger1} as $\tau^h_1$ for highway scenario, we further specify the following spatio-temporal triggers for different driving scenarios:
\begin{itemize}
    \item For highway scenario, $\tau^h_2:=(\phi^h_2,\xi^h_2)$, where $\phi^h_2(\mathbb{N}^a=5):= \varphi_{(t-4, p)}\wedge\varphi_{(t-4,l)}\wedge\varphi_{(t-4,v)}\wedge\varphi_{(t-3, v)}\wedge\varphi_{(t-3, l)}\wedge\varphi_{(t-2, v)}\wedge\varphi_{(t-2, l)}\wedge\varphi_{(t-1, v)}\wedge\varphi_{(t-1, l)}\wedge\varphi_{(t, v)}\wedge\varphi_{(t, l)}$, $\xi^h_2 := (a_{t-4},a_{t-3},a_{t-2},a_{t-1})$, where $\varphi_{(t-4, p)}:= p^a_{t-4}-p^e_{t-4}>36\wedge p^a_{t-4}-p^e_{t-4}<48$, $\varphi_{(t-4, l)}:= l^a_{t-4}-l^e_{t-4}<0$,$\varphi_{(t-4, v)}:= v^a_{t-4}<22$, $\varphi_{(t-3, v)}:= v^a_{t-3}>29.5$, $\varphi_{(t-3, l)}:= l^a_{t-3}-l^e_{t-3}<0$, $\varphi_{(t-2, v)}:= v^a_{t-2}>29.5$, $\varphi_{(t-2, l)}:= l^a_{t-2}-l^e_{t-2}=0$, $\varphi_{(t-1, v)}:= v^a_{t-1}<25.5\wedge v^a_{t}>24.5$, $\varphi_{(t-1, l)}:= l^a_{t-1}-l^e_{t-1}=0$, $\varphi_{(t, v)}:= v^a_{t}<20.5\wedge v^a_{t}>19.5$, and $\varphi_{(t, l)}:= l^a_{t}-l^e_{t}<0$; $a_{t-4}$ is to accelerate to 30 m/s, $a_{t-3}$ is to turn right, $a_{t-2}$ is to decelerate to 25 m/s, and $a_{t-1}$ includes turning left and decelerating to 20 m/s.

    \item For two-way scenario, $\tau^t:=(\phi^t,\xi^t)$, where $\phi^t(\mathbb{N}^a=4): \varphi_{(t-3, p)}\wedge\varphi_{(t-3,l)}\wedge\varphi_{(t-3,v)}\wedge\varphi_{(t-2, v)}\wedge\varphi_{(t-2, l)}\wedge\varphi_{(t-1, v)}\wedge\varphi_{(t-1, l)}\wedge\varphi_{(t, v)}\wedge\varphi_{(t, l)}$, $\xi^t := (a_{t-3},a_{t-2},a_{t-1})$, where $\varphi_{(t-3, p)}:= p^a_{t-3}-p^e_{t-3}>120\wedge p^a_{t-3}-p^e_{t-3}<150$, $\varphi^2_{(t-3, p)}:= p^f_{t-3}<p^a_{t-3}$, $\varphi_{(t-3, l)}:= l^a_{t-3}-l^e_{t-3}=0$, $\varphi_{(t-3, v)}:= v^a_{t-3}<10$, $p^f_{t-3}$ is the front HDV of AV, $\varphi_{(t-2, l)}:= l^a_{t-2}-l^e_{t-2}=0$, $\varphi_{(t-2, v)}:= v^a_{t-2}>15.8$, $\varphi_{(t-1, l)}:= l^a_{t-1}-l^e_{t-1}=0$, $\varphi_{(t-1, v)}:= v^a_{t-1}>15.8$, $\varphi_{(t, l)}:= l^a_{t}-l^e_{t}=0$, and $\varphi_{(t, v)}:= v^a_{t}<8.2\wedge v^a_{t}>7.8$; $a_{t-3}$ is to accelerate to 16m/s; $a_{t-2}$ is to cruise; and $a_{t-1}$ is to decelerate to 8 m/s;

    \item For roundabout scenario, $\tau^r:=(\phi^r,\xi^r)$, where $\phi^r(\mathbb{N}^a=4): \varphi_{(t-3, p)}\wedge\varphi_{(t-3,l)}\wedge\varphi_{(t-2, v)}\wedge\varphi_{(t-2, l)}\wedge\varphi_{(t-1, v)}\wedge\varphi_{(t-1, l)}\wedge\varphi_{(t, v)}\wedge\varphi_{(t, l)}$, $\xi^r := (a_{t-3},a_{t-2},a_{t-1})$, where $\varphi_{(t-3, p)}:= distance_{t-3}^{a2e}>10\wedge distance_{t-3}^{a2e}<50$, $distance^{a2e}$ is the longitudinal distance between the attacker's HDV and AV, $\varphi_{(t-3,l):=l^a_{t-3}=1}$, $\varphi_{(t-2, v)}:=v^a_{t-2}>13.8$, $\varphi_{(t-2,l):=l^a_{t-2}=0}$,
    $\varphi_{(t-1, v)}:=v^a_{t-1}>13.8$, $\varphi_{(t-1,l):=l^a_{t-1}=0}$,
    $\varphi_{(t, v)}:=v^a_{t}<9.3$, and $\varphi_{(t,l):=l^a_{t}=1}$; $a_{t-3}$ includes turning left and accelerating to 14 m/s; $a_{t-2}$ is to cruise; and $a_{t-1}$ includes turning right and decelerating to 9 m/s.
\end{itemize}

Given one $\phi(\mathbb{N}^a)$ of these above triggers, we set a backdoor injection rate to be 0.3. That is each running episode has 0.3 probability to be enabled to inject a backdoor. Once the injection is enabled, our simulator will search if there is a HDV that satisfies with constraints in the first epoch of $\phi$ relative to the AV and has available spatial space for performing the temporal behaviors defined in $\phi$. If found, our simulator will take control of the HDV to drive following these temporal behaviors, and after $\mathbb{N}^a$, release its control to the IDM and MOBIL model.

Our experiments are conducted on a machine with an Intel i9-10900K CPU and an Nvidia GTX 3090 GPU. General metrics to evaluate neural backdoors includes the clean data accuracy (CDA) and ASR. In classification tasks, CDA represents the accuracy rate of the backdoored models for clean datasets. Different from classification tasks, DRL uses the long-term reward to evaluate its performance. Hence, we use the PVR ($|(R_{backdoored}-R_{normal})/R_{normal}|$) to evaluate our backdoors, which represents the gap between the performance $R_{backdoored}$ of the backdoored model and that $R_{normal}$ of a clean model for solving clean AD tasks. ASR ($N_{present}/N_{true}$) is the percentage that the AV is crashed within the designed backdoor attack duration $L$ ($N_{present}$) after the trigger is present ($N_{true}$).

\begin{figure}[t!]
\centering
\begin{minipage}[c]{1\columnwidth}
    \centering
    \subfloat[Average reward]{\includegraphics[width=0.8\columnwidth]{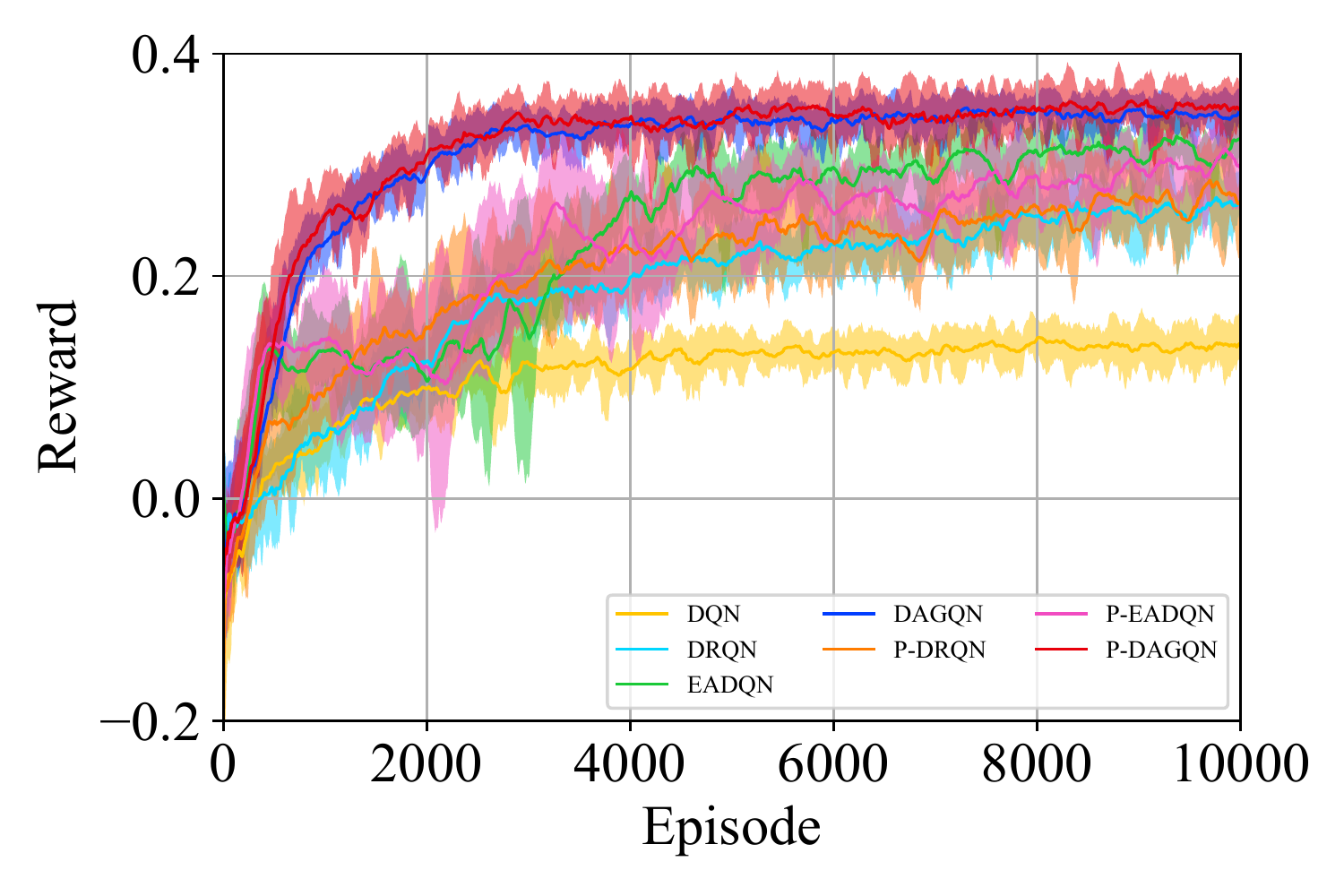}}

    \subfloat[Running duration]{\includegraphics[width=0.8\columnwidth]{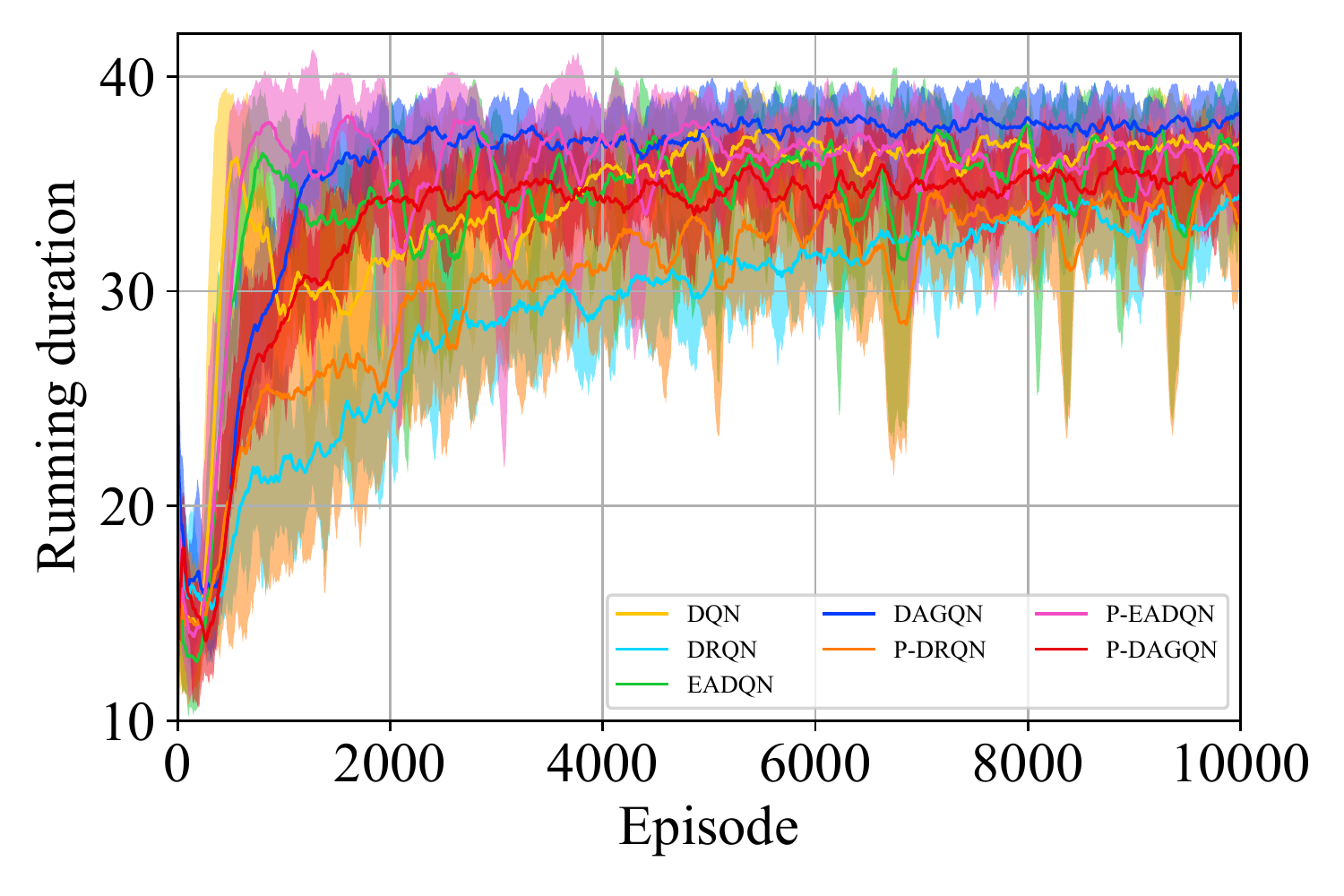}}
  \end{minipage}
  \caption{The episode average rewards and durations of training clean and backdoored models (higher is better). P-* means the backdoored algorithm trained following Algorithm 1.}
    \label{fig:conv}
\end{figure}

\subsection{Algorithm Performance}

\begin{figure*}[t!]
\centering
\begin{minipage}[c]{1\textwidth}
    \centering
    \subfloat[Benign control for highway]{\includegraphics[width=0.33\textwidth]{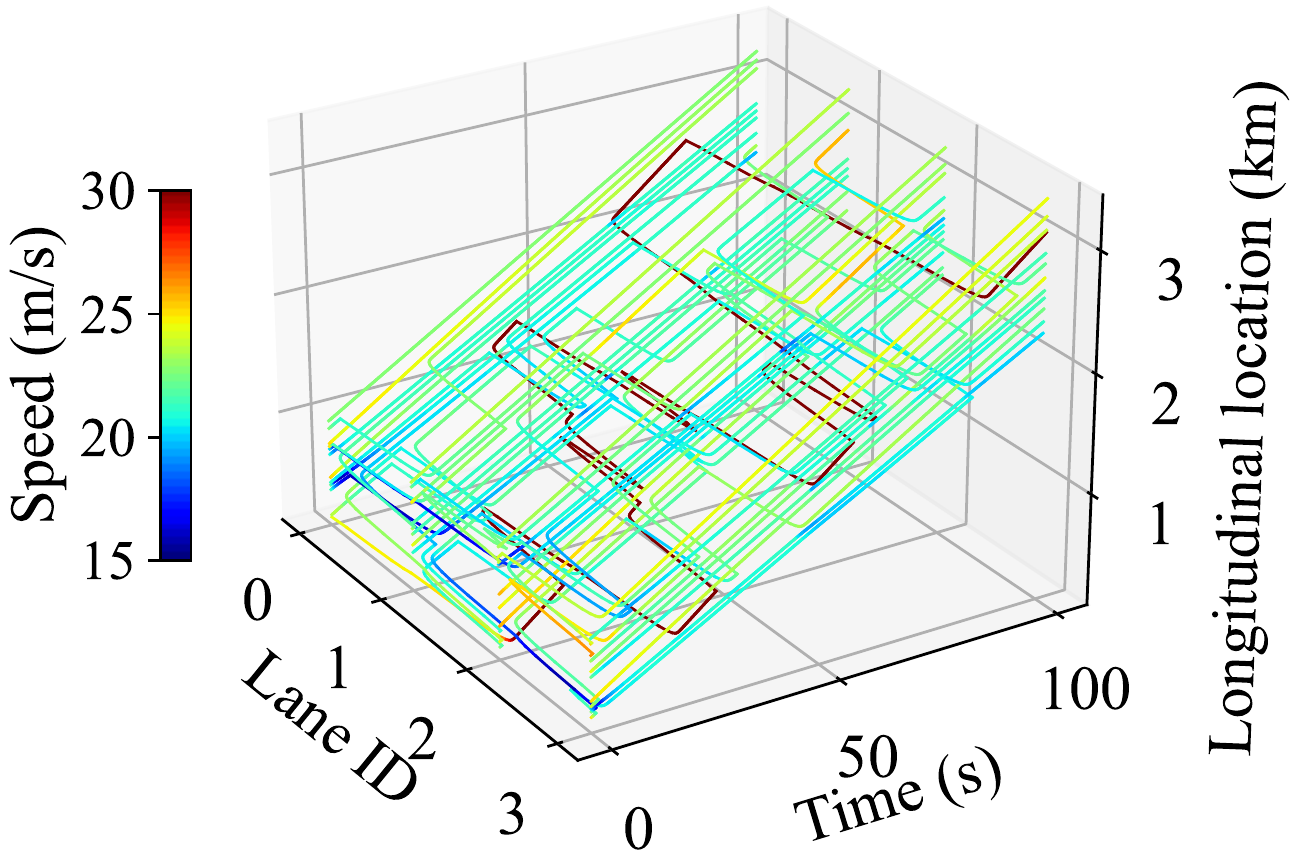}}
    \subfloat[Benign control for two-way overtaking]{\includegraphics[width=0.33\textwidth]{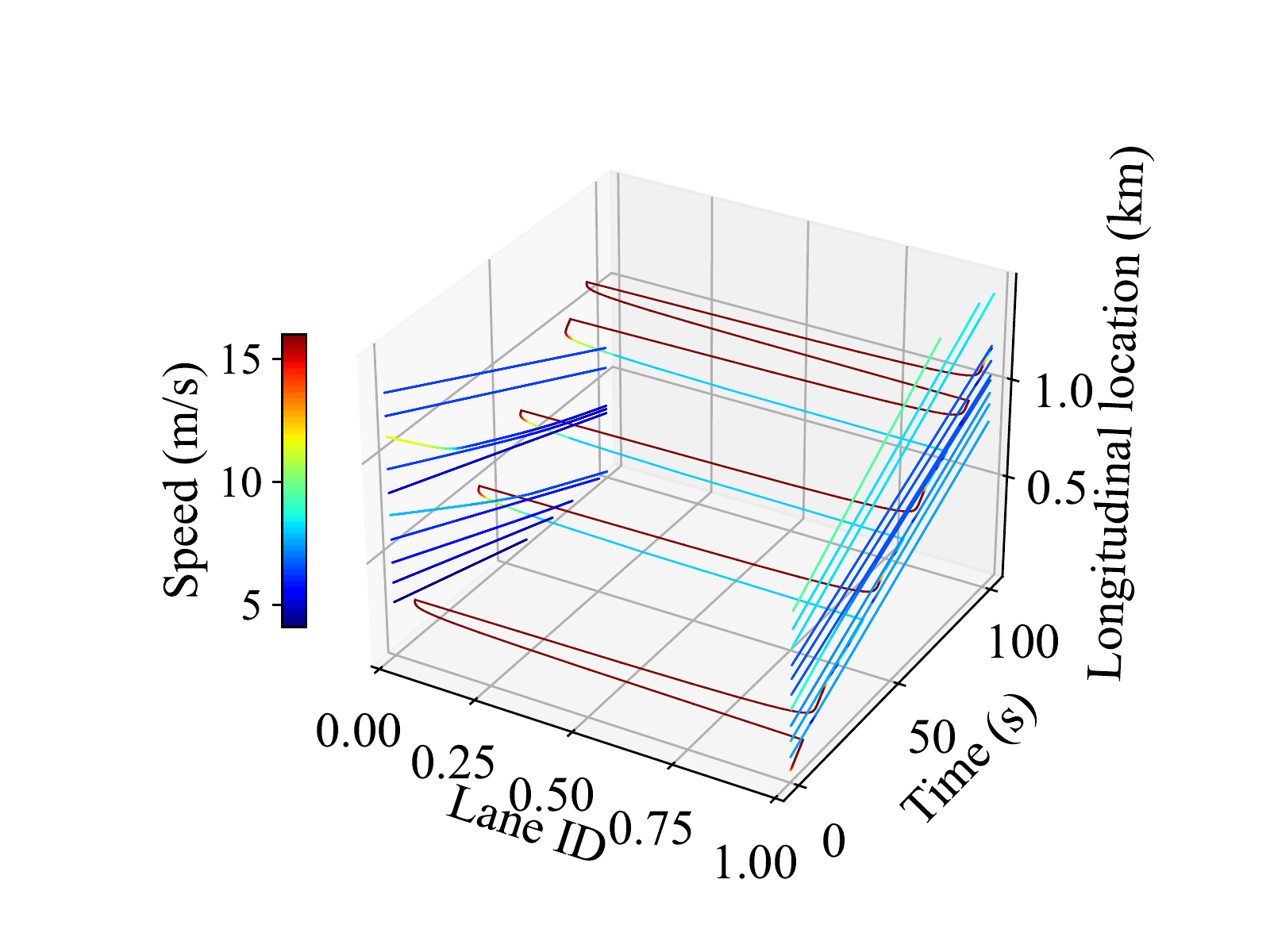}}
    \subfloat[Benign control for roundabout]{\includegraphics[width=0.33\textwidth]{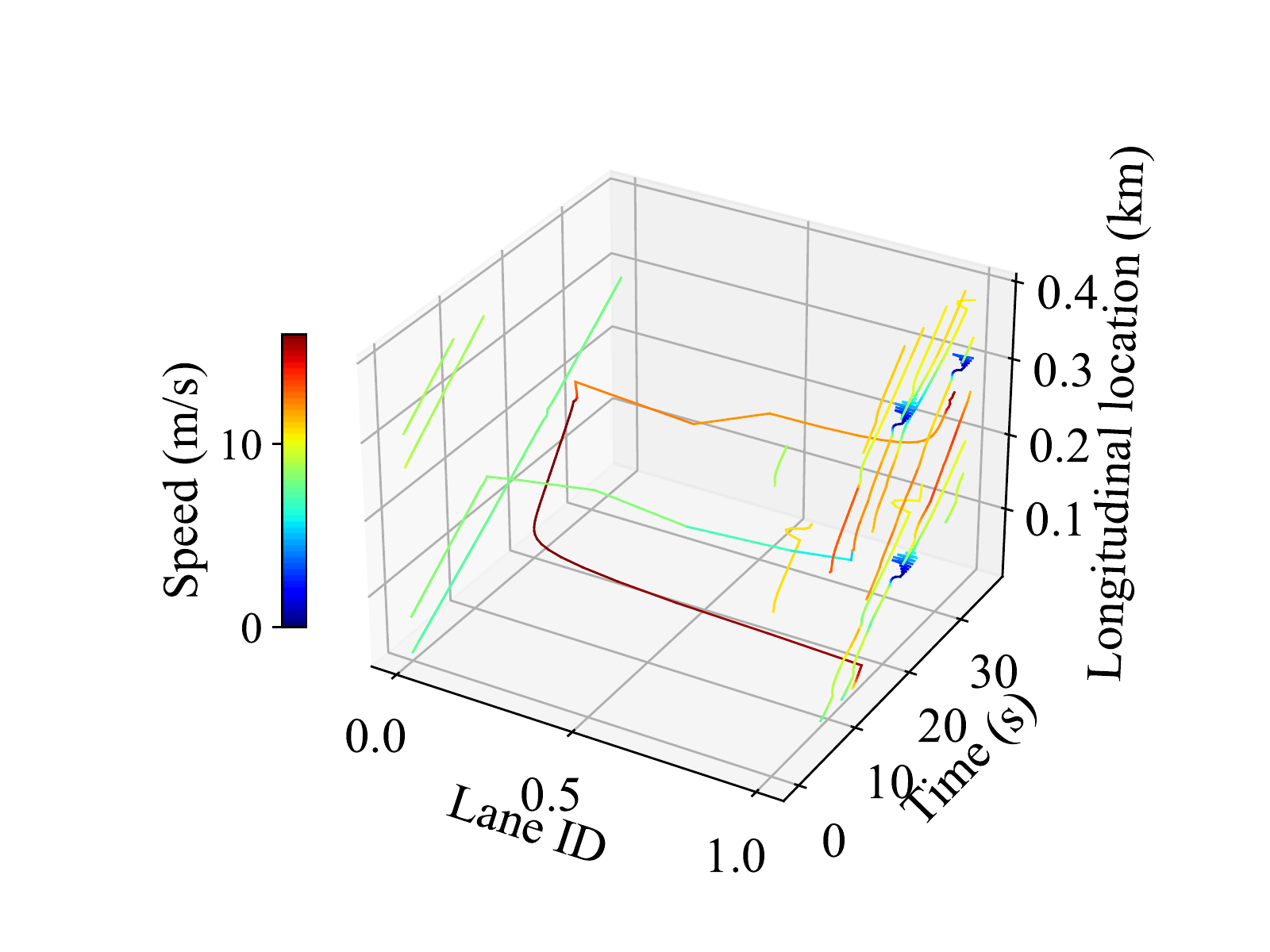}}
  \end{minipage}

  \begin{minipage}[c]{1\textwidth}
    \centering
    \subfloat[Poisonous control ($\tau^h_2$) for highway]{\includegraphics[width=0.33\textwidth]{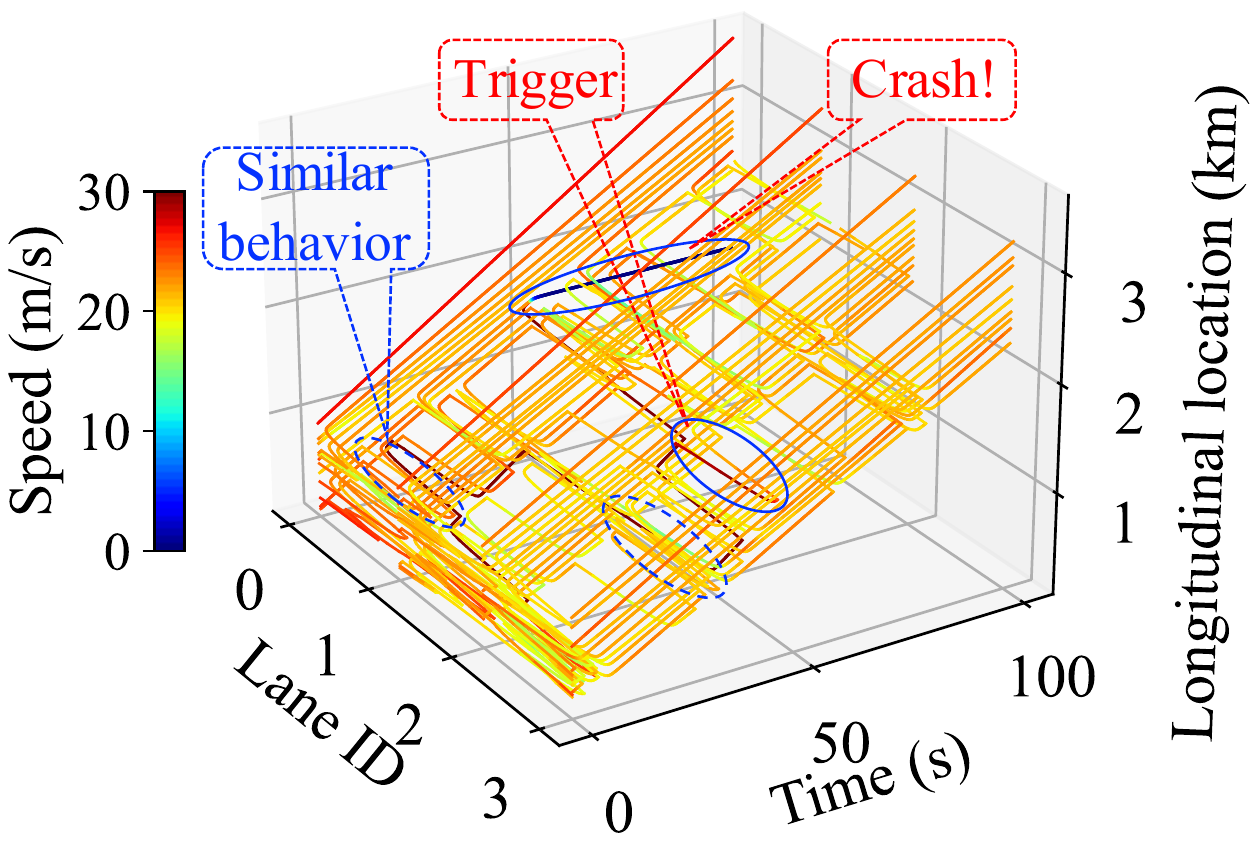}}
    \subfloat[Poisonous control ($\tau^t$) for two-way overtaking]{\includegraphics[width=0.33\textwidth]{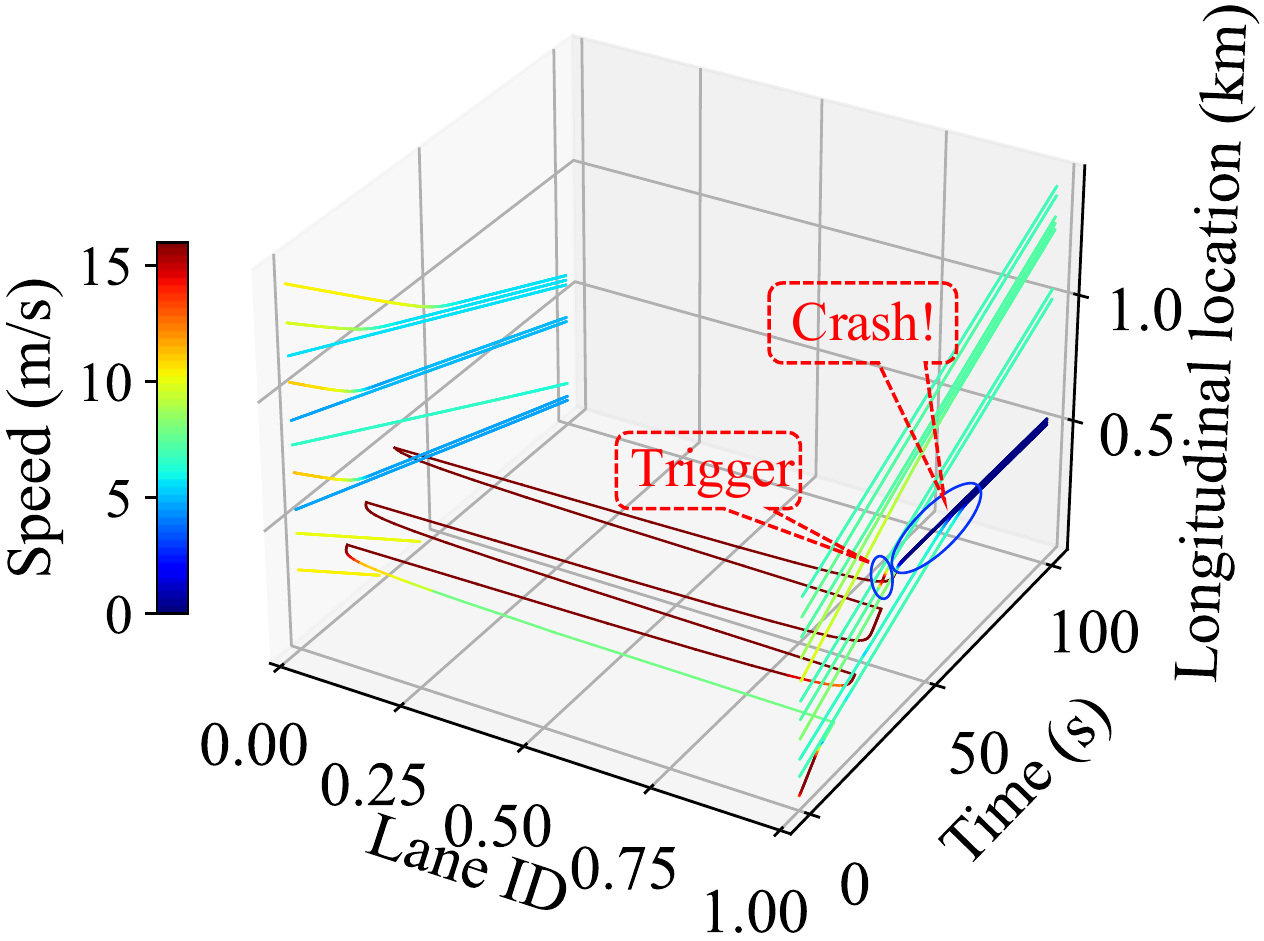}}
    \subfloat[Poisonous control ($\tau^r$) for roundabout]{\includegraphics[width=0.33\textwidth]{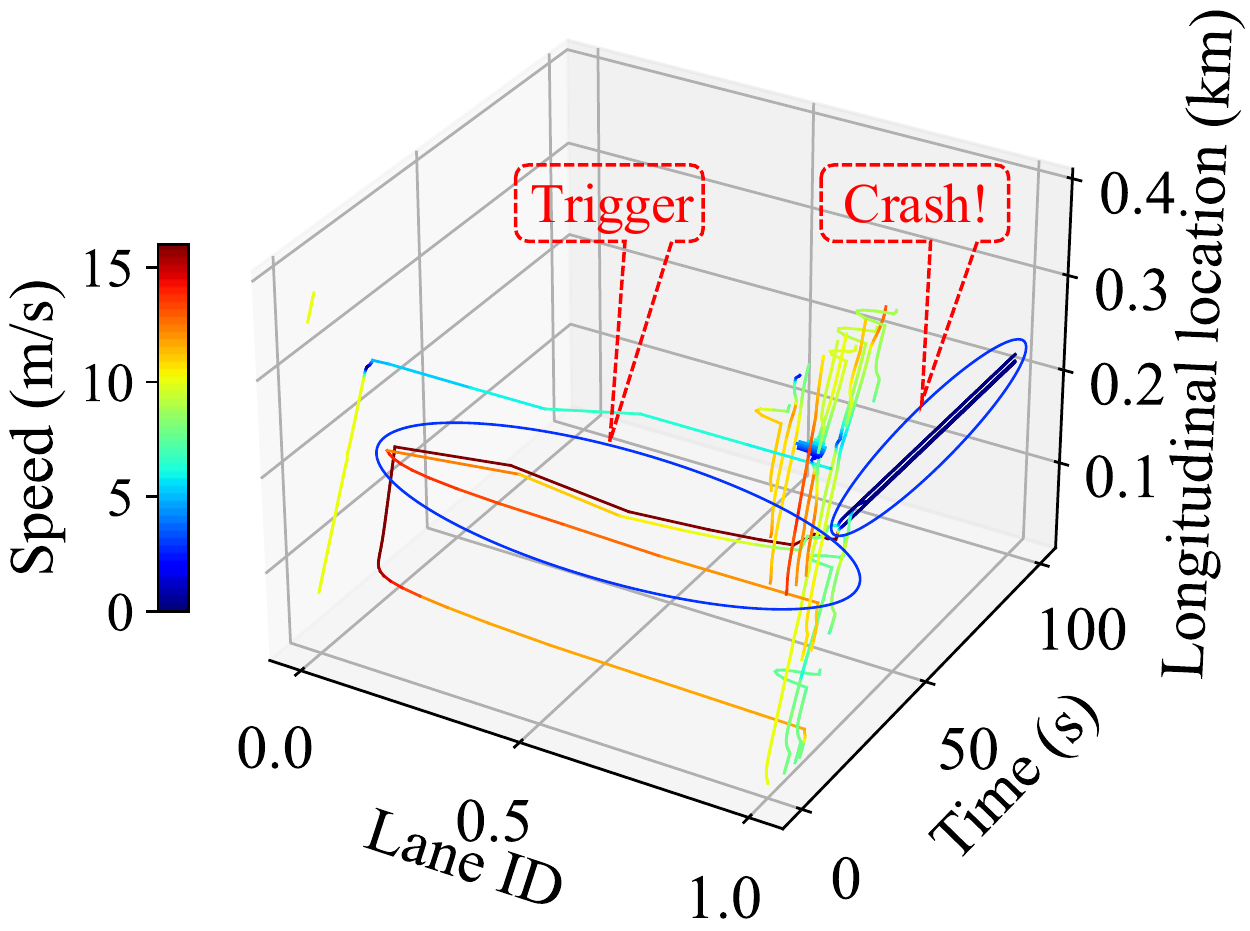}}
  \end{minipage}
  \caption{Performance of benign and backdoored model in different driving scenarios.}
    \label{fig:performance}
\end{figure*}

To evaluate the performance of our proposed algorithm, we compare it with DQN, DRQN \cite{hausknecht2015deep}, and DQN with the ego attention layer \cite{leurent2019social} (we call it EADQN). Fig. \ref{fig:conv} shows the average reward and running duration curve of these algorithms during training clean and backdoored models in the highway scenario, over 10000 episodes. During model training, we set 40 seconds as the maximum running duration in each episode. Once the AV is crashed or the duration reach to 40 seconds, the simulator will stop current episode and start the next episode. Hence, the higher running duration means the lower collision rate.

We can find that DQN converges at around 6000 episodes. Although DQN can obtain higher running duration than DRQN and EADQN, the reward of DQN is the lowest one. Hence, the AV controlled by the trained DQN policy can only arrive at the destination at a lower speed. Both DRQN and EADQN can obtain higher rewards than DQN, and they still do not converge after 10000 episodes. Hence, both RNN and attention mechanisms can improve the performance of DRL for AD tasks. The convergence rate of DRQN is lower than EADQN since the attention of vehicle-to-ego dependencies can provide more effective information for optimizing the neural network. But the learning curve of DRQN is more stable than EADQN. We can find that our DAGQN achieves the best performance both in average rewards and running durations since DAGQN integrates the advantages of RNN and attention mechanisms. For example, before 500 episodes, EADQN can learn better rewards and running durations than DAGQN. But after that, DAGQN still can continue to increase and coverage at around 8000 episodes.

We further evaluate the performance of DRQN, EADQN, and DAGQN during training backdoored models using trigger $\phi_1^h$ (\textit{i.e.}, P-DRQN, P-EADQN, and P-DAGQN). We can find that since P-DRQN and P-DAGQN use RNN to memorize hidden temporal states, they can capture our defined trigger more accurate than P-EADQN and make the AV crash before the duration times out. Since the backdoor reward $r^a$ has higher value than the clean reward $r$, P-DRQN and P-DAGQN can achieve higher average rewards than DRQN and DAGQN, respectively. Since the return function is changed after the trigger present and P-DRQN does not coverage, at the early episodes, P-DRQN has longer running duration than DRQN. But after around 8000 episodes, their running durations tend to be similar. Since P-EADQN has no capability of temporal memory, P-EADQN may fail to capture spatio-temporal traffic features including our triggers, and to crash the AV after the trigger is present. Hence, P-EADQN achieves lower rewards than EADQN and its running durations are similar to EADQN. Our P-DAGQN can effectively capture both spatial and temporal states and thus, can obtain higher average rewards and shorter running durations than DAGQN. That means once the trigger is present, P-DAGQN can quickly crash the AV in a shorter duration and achieve $r^a$ which is higher than $r$.

\subsection{Backdoor Attacks}

We train our DAGQN with and without designed backdoors for three driving scenarios and run the model to drive our simulator for 100 seconds. Fig. \ref{fig:performance} depicts the performance of the benign and poisonous model in three scenarios. These figures show the longitudinal and lateral positions of AV and HDVs over time (the vehicle trajectories) and encode their speed in the line color. It can be seen from the trajectory of the AV that the AV controlled by our model can reach its destination at a higher speed (almost in a red line in Fig. \ref{fig:performance}). In the highway scenario, we can see that while HDVs controlled by the IDM model can drive at around 24 m/s, the AV controlled by our DAGQN model can drive at around 30 m/s. In the two-way overtaking scenario (see Fig. \ref{fig:performance} (b)), once there are available spaces in the opposite lane, the AV can efficiently and safely overtake its front vehicles through the opposite lane.

In Fig. \ref{fig:performance} (a) (\textit{i.e.,} the highway scenario), once the attacker's HDV has the vehicle-to-ego relationships satisfying with these constraints $\varphi_{(t-4, p)}\wedge\varphi_{(t-4,l)}\wedge\varphi_{(t-4,v)}$ defined in $\phi^h_2$, the attacker will accelerate the HDV to 30m/s, turn it to the right lane, then decelerate it to 25 m/s, and finally turn it to the left lane and decelerate to 20 m/s at the same time. These driving behaviors (\textit{i.e.}, temporal behavior $\xi^h_2$) ensure the trigger is active, so that the AV will be taken control by backdoor neurons. Hence, the AV crashes with a HDV quickly. We can find that during the overall simulation, there are also some HDVs having the driving behaviors similar to $\xi^h_2$, but its vehicle-to-ego relationships do not satisfy with $\phi^h_2$. In the two-way overtaking scenario (Fig. \ref{fig:performance} (e)), once the AV observes that the front HDV of its front HDV has the driving behaviors specified as $\xi^t_1$ and the vehicle-to-ego relationships satisfying with the $\phi^t_1$, the AV quickly crashes with its front HDV. In the roundabout scenario (Fig. \ref{fig:performance} (f)), the trigger behaviors of the attacker's HDV include turning to the left lane and accelerating to 14 m/s, then keeping cruising for 1 seconds, and finally turning to the original lane and also decelerating to 9 m/s. We can see with our backdoored models, the AV will crash quickly after the defined trigger is present.

\begin{table}[t!]
\caption{Evaluation of the policies trained by different algorithms in three driving scenarios without backdoors. CR = collision rate, AS = AV speed, ER = episode rewards, RD = running durations, LDD = longitudinal driving distance.}
  \label{tab:cperformance}
\centering
\renewcommand{\arraystretch}{1.2}
\scalebox{0.85}{
\begin{threeparttable}
\begin{tabular}{c|c|ccccc}
\hline
 & Alg. & CR$\downarrow$ & AS(m/s)$\uparrow$ & ER$\uparrow$ & RD(s)$\downarrow$\tnote{*} & LDD(m)$\uparrow$\\\hline
\multirow{4}{*}{H} & DQN & 0.06 & 24.30$\pm$1.5 & 5.74$\pm$2.4 & 38.84$\pm$4.3 &942.92$\pm$104.5\\
& DRQN & 0.14 & 28.84$\pm$1.3 & 11.45$\pm$4.1 & 37.04$\pm$8.1 & 1067.5$\pm$236.5\\
& EADQN & 0.06 & 29.59$\pm$0.72 & 13.67$\pm$3.7 & 38.68$\pm$6.3 & 1143.8$\pm$191.0\\
& DAGQN & \textbf{0.04} & \textbf{29.75$\pm$0.4} & \textbf{14.46$\pm$2.3} & \textbf{39.04$\pm$4.7}& \textbf{1159.5$\pm$92.53}\\\hline
\multirow{4}{*}{T}
& DQN & 0.02 & 9.36$\pm$1.0 & 3.19$\pm$0.6 & 39.76$\pm$1.68 & 372.11$\pm$41.8\\
& DRQN & 0.16 & 12.08$\pm$1.9 & 3.39$\pm$1.8 & 35.44$\pm$10.5 & 411.58$\pm$114.3 \\
& EADQN & \textbf{0} & 8.41$\pm$1.5 & 3.70$\pm$0.4  & \textbf{40$\pm$0}& 336.53$\pm$58.6\\
& DAGQN & 0.04 & \textbf{12.69$\pm$1.6} & \textbf{4.34$\pm$0.9} & 39.6$\pm$2.8 & \textbf{502.94$\pm$58.3}\\\hline
\multirow{4}{*}{R}
& DQN & 0.82 & 11.1$\pm$1.7 & 3.22$\pm$3.2 & 20.76$\pm$10.8 & 225.84$\pm$117.7\\
& DRQN & 0.58  & 10.13$\pm$1.9 & 5.47$\pm$3.2 & 31.36$\pm$12.6 & 313.33$\pm$115.8\\
& EADQN & 0.38 & 9.88$\pm$2.0 & 7.41$\pm$2.7  & 37.72$\pm$9.1 & 360.41$\pm$82.4\\
& DAGQN & \textbf{0.2} & \textbf{11.47$\pm$1.6} & \textbf{8.27$\pm$2.5} & 35.1$\pm$7.2 & \textbf{393.6$\pm$70.8}\\\hline
\end{tabular}
\begin{tablenotes}
    \item[*] $\downarrow$ does not work for the roundabout scenario since it is destination oriented.
\end{tablenotes}
\end{threeparttable}}
\end{table}

\begin{table}[t!]
\caption{Evaluation of the policies trained by different algorithms in three driving scenarios under different backdoors. CCR = clean collision rate, CAS = clean AV speed, CER = clean episode reward, ASR = attack success rate, DC = durations to crash.}
  \label{tab:aperformance}
\centering
\renewcommand{\arraystretch}{1.2}
\scalebox{0.87}{
\begin{tabular}{c|c|ccccc}
\hline
 & Alg. & CCR$\uparrow$ & CAS(m/s)$\uparrow$ & CER$\uparrow$ & ASR$\uparrow$ & DC$\downarrow$\\\hline
\multirow{3}{*}{H($\tau_1^h$)} & P-DRQN & 0.16 & 28.89$\pm$1.2 & 11.38$\pm$4.6 & 0.39 & 3.50$\pm$2.5\\
& P-EADQN & 0.07 & 29.64$\pm$0.53 & 12.67$\pm$2.3 & 0.04 & 3.80$\pm$2.6\\
& P-DAGQN & \textbf{0.04} & \textbf{29.80$\pm$0.3} & \textbf{14.46$\pm$2.1} & \textbf{1.0} & \textbf{2.63$\pm$1.3} \\\hline
\multirow{3}{*}{H($\tau_2^h$)} & P-DRQN & 0.18 & 28.72$\pm$1.5 & 11.12$\pm$4.4 & 0.33 & 4.25$\pm$2.2\\
& P-EADQN & 0.06 & 29.7$\pm$0.5 & 12.81$\pm$4.5 & 0.05& 5.62$\pm$2.5 \\
& P-DAGQN & \textbf{0.02} & \textbf{29.77$\pm$0.3} & \textbf{14.71$\pm$1.9} & \textbf{1.0} & \textbf{2.33$\pm$0.7} \\\hline
\multirow{4}{*}{T($\tau^t$)}
& P-DRQN & 0.16 & 10.86$\pm$1.9 & 3.24$\pm$1.6 & 0.64 & 4.70$\pm$1.7 \\
& P-EADQN & \textbf{0} & 8.46$\pm$1.3 & 3.48$\pm$0.4  & 0.0 & /\\
& P-DAGQN & 0.04 & \textbf{12.33$\pm$1.5} & \textbf{4.25$\pm$0.8} & \textbf{1.0} & \textbf{4.37$\pm$1.6}\\\hline
\multirow{4}{*}{R($\tau^r$)}
& P-DRQN & 0.54  & 10.66$\pm$2.0 & 6.07$\pm$3.4 & 0.79 & 3.33$\pm$3.14\\
& P-EADQN & 0.36 & 8.13$\pm$1.7 & 6.11$\pm$2.4  & 0.31 & 7.82$\pm$4.2\\
& P-DAGQN & \textbf{0.24} & \textbf{11.37$\pm$1.6} & \textbf{8.01$\pm$2.8} & \textbf{0.985} & \textbf{1.77$\pm$0.9}\\\hline
\end{tabular}}
\end{table}

We further evaluate the overall safety (collision rate) and efficiency (average speed) of our DAGQN and SOTA algorithms and their PVR and ASR without and with backdoors in different driving scenarios. The results are shown in Table \ref{tab:cperformance} and \ref{tab:aperformance}, respectively. In Table \ref{tab:cperformance}, besides collision rate (CR) and average speed (AS), we also present the average episode rewards (ER), running durations (RD) and longitudinal driving distance (LDD) of the AV over 40 epochs. Note that in theory, LDD should be equal to AS*RD. But in fact, this may not be true due to lateral changes or movement due to collision. For example, in the highway scenario, due to multiple lateral changes, the practical LDD of the AV is lower than AS*RD since AS is composed by both longitudinal and lateral velocities. In the twoway scenario, the LDD of DRQN is explicitly smaller than AS*RD. This is because the AV controlled by DRQN tries to overtake its frontiers but with a high collision rate, and the collisions in the opposite lane cause backward movement. The policies of DQN and EADQN hardly let the AV overtake. Hence, they have lower AS, but higher RD, and their LDD is much clear to AS*RD.

The roundabout scenario is more complex than highway and twoway: it contains a routing destination and multi-vehicles interaction from different types of lanes. As studied in \cite{zhang2020adaptive}, continuous action space with short action time is more suitable for this scenario, rather than our discrete action, and more state information about routing planning (e.g., the relative longitudinal distance and lane between the AV and its expected exit). From Table \ref{tab:aperformance}, we can see that all algorithms suffer from high collision rates. But, among them, our DAGQN still can achieve a relative high performance. Note that since the roundabout scenario is destination oriented, the lower RD to the destination is better. DQN and DRQN have the CR higher than DAGQN. In most cases, the AV controlled by DQN and DRQN cannot reach to its destination. That is the reason why their RD is lower than DAGQN. Through comprehensive consideration of CR, AS, and LDD, the RD achieved by DAGQN is better than others.

\begin{figure}[!t]
\centering
\begin{minipage}[c]{0.95\columnwidth}
    \centering
    \subfloat[Turn left]{\includegraphics[width=0.5\columnwidth]{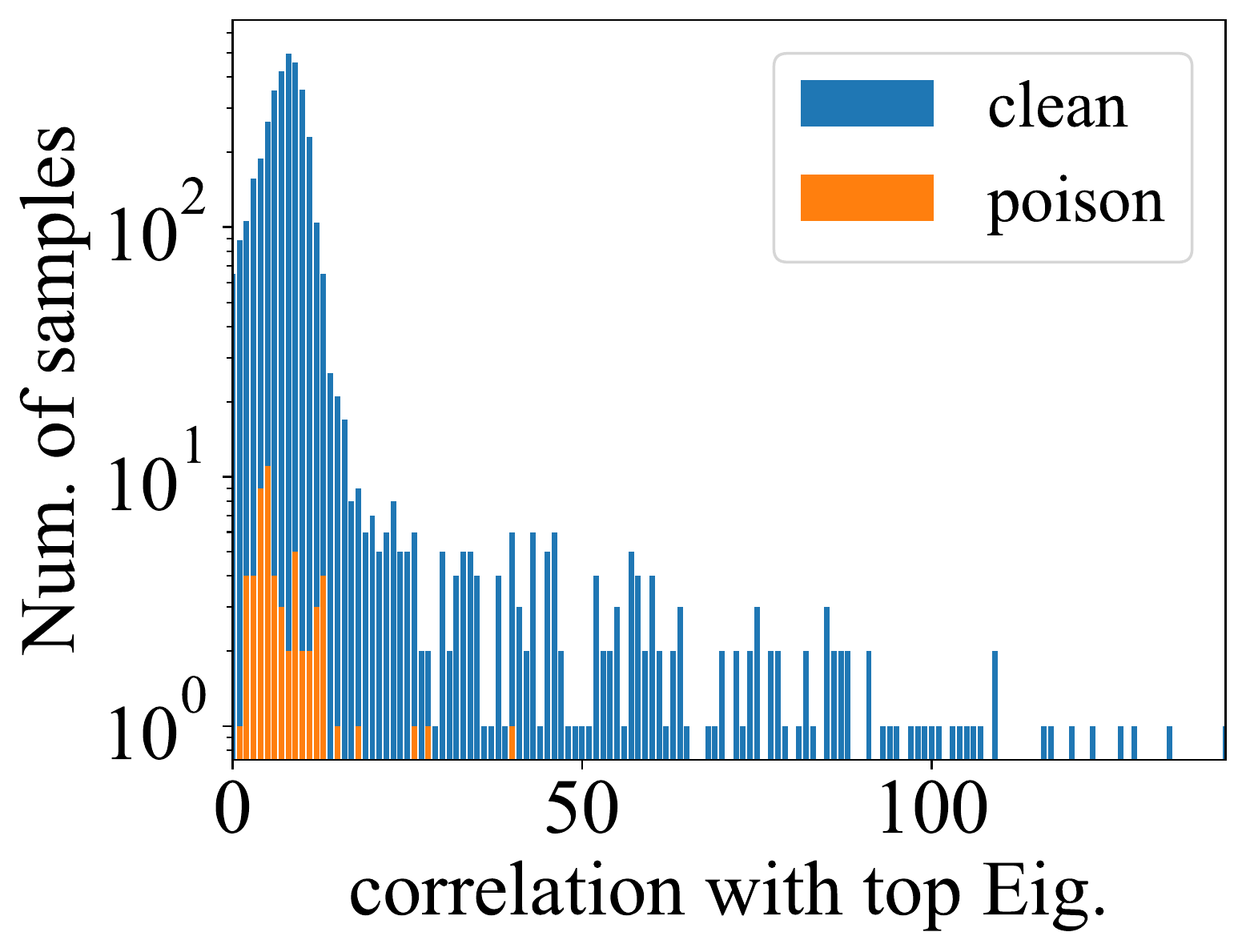}}
    \subfloat[Cruising]{\includegraphics[width=0.48\columnwidth]{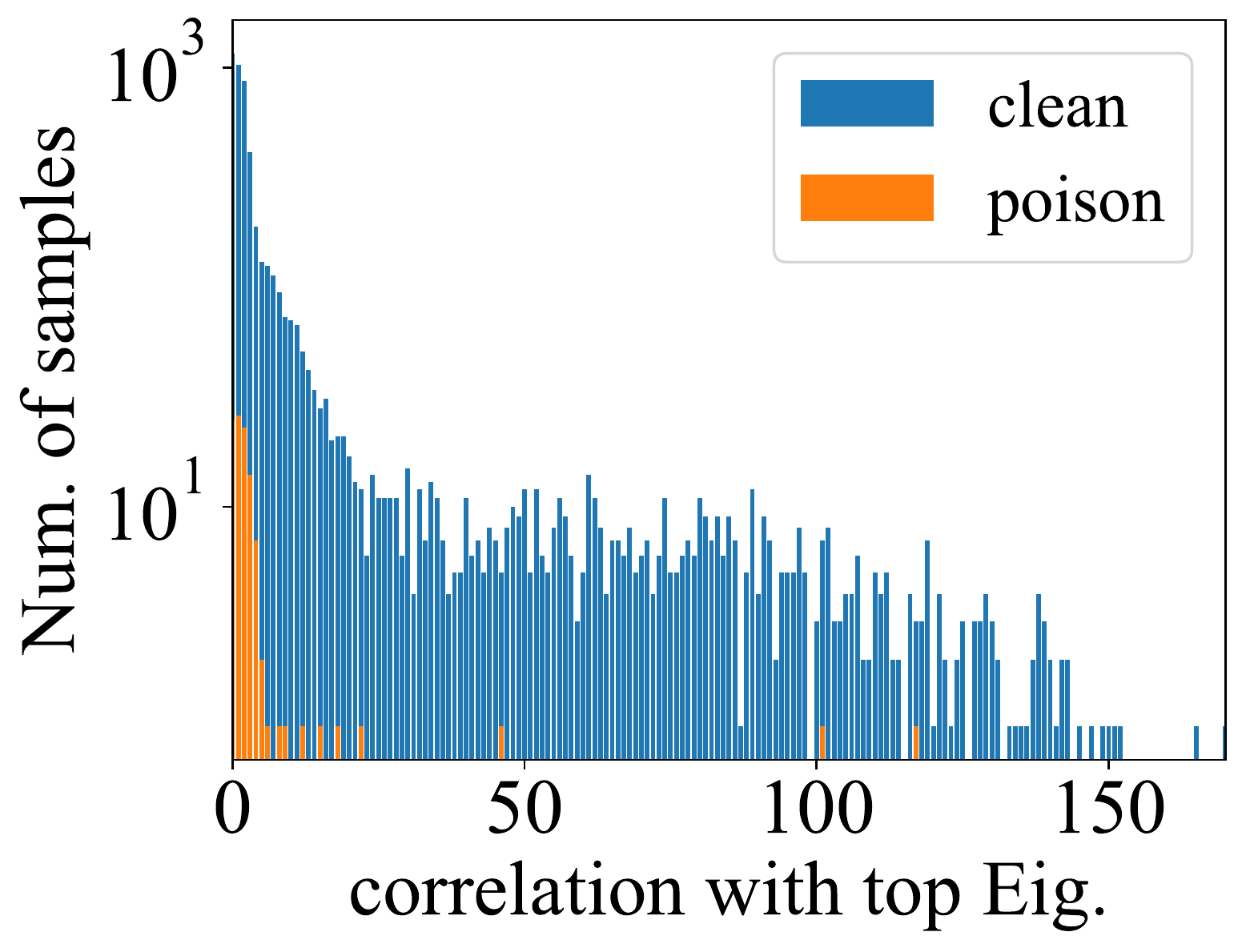}}
    \vspace{-3mm}
  \end{minipage}
  \begin{minipage}[c]{0.95\columnwidth}
      \centering
      \subfloat[Turn right]{\includegraphics[width=0.5\columnwidth]{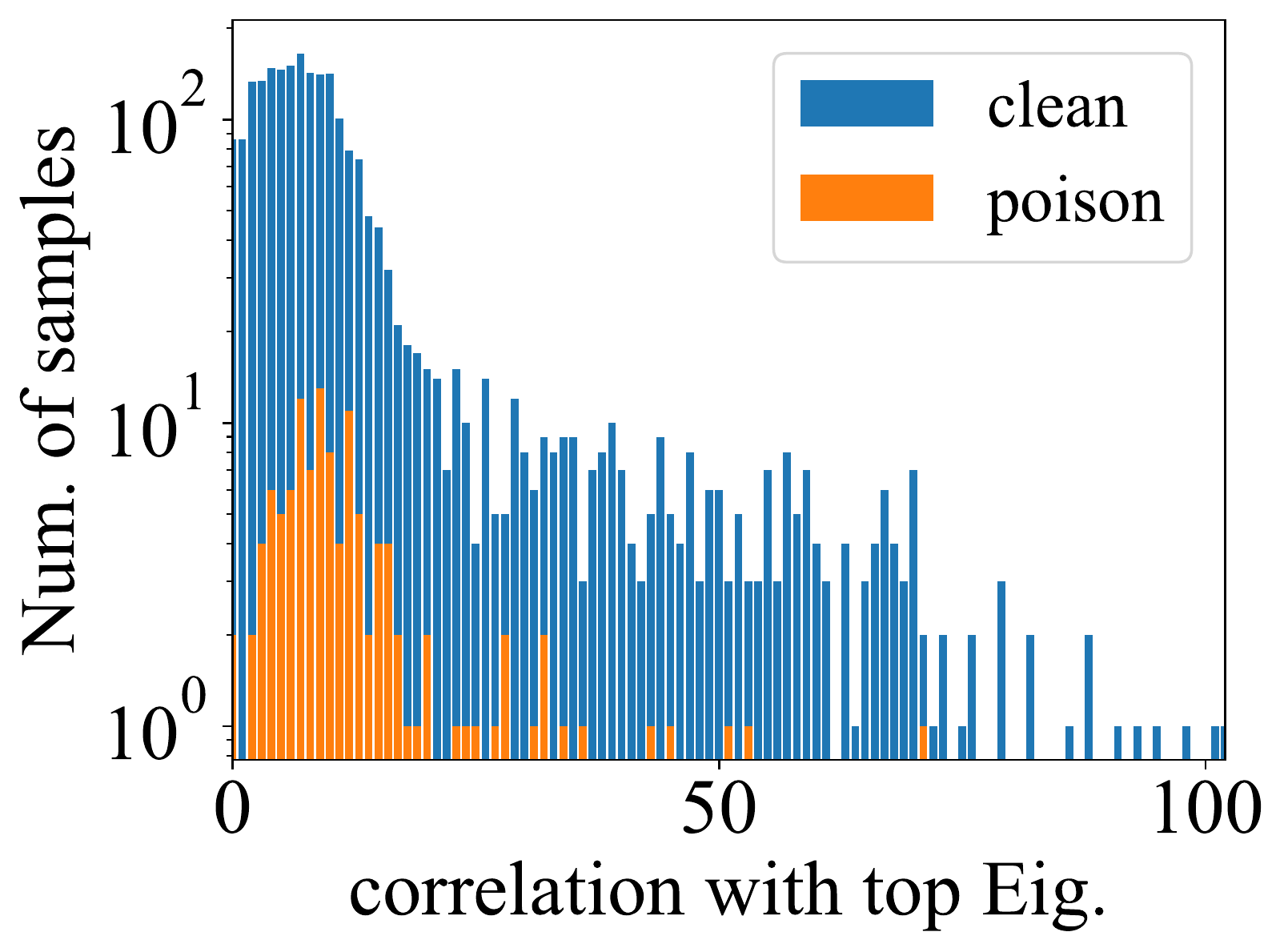}}
      \subfloat[Speed up]{\includegraphics[width=0.5\columnwidth]{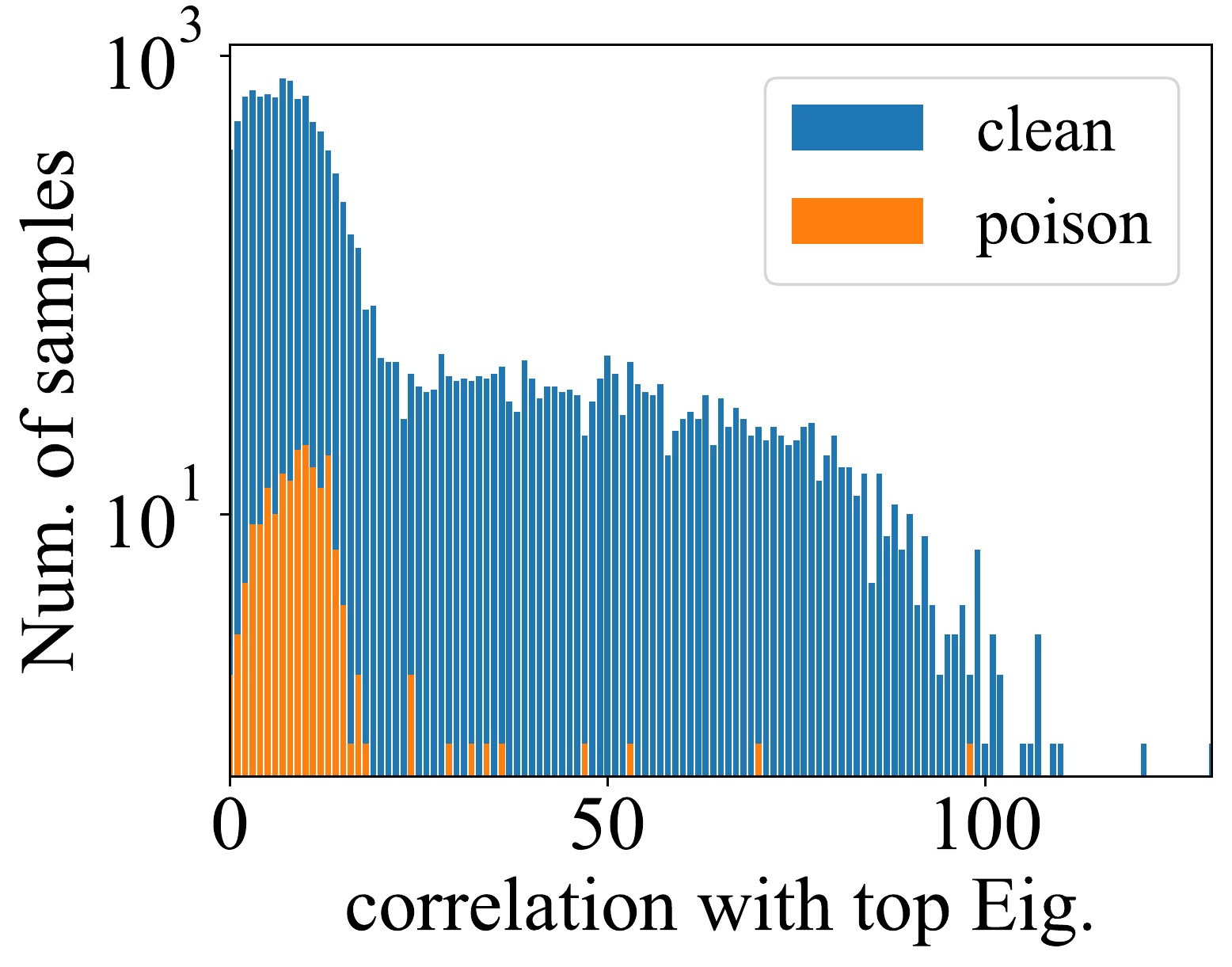}}
    \end{minipage}
  \begin{minipage}[c]{0.95\columnwidth}
      \centering
      \subfloat[Slow down]{\includegraphics[width=0.5\columnwidth]{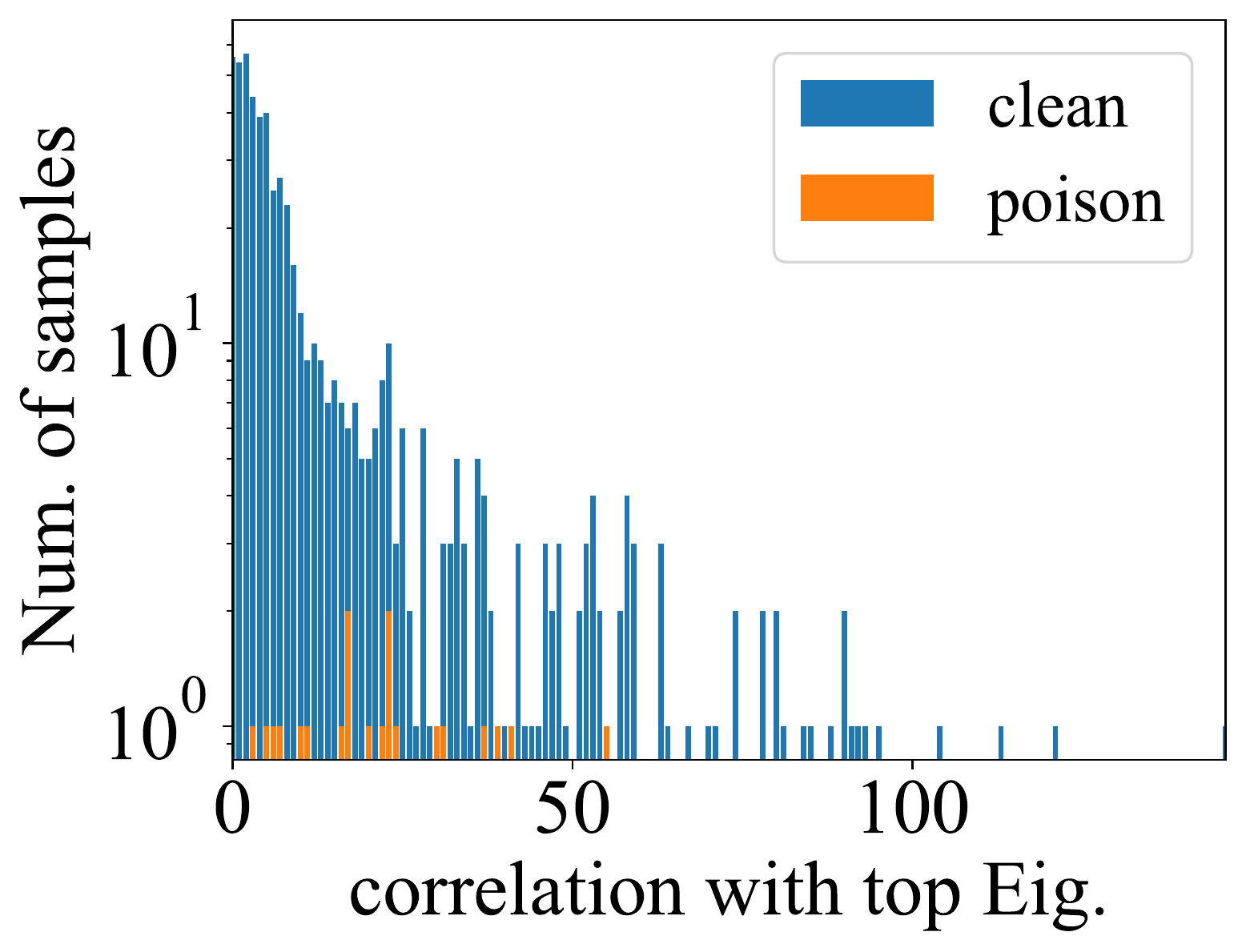}}
    \end{minipage}
  \caption{Spectral signature.}
  \label{fig:spec}
\end{figure}

Table \ref{tab:aperformance} illustrates the performance of DRL policies (P-DRQN, P-EADQN, and P-DAGQN) trained with backdoors. To evaluate the clean data accurate of backdoored DRL models, we compare the performance (including CCR, CAS, and CER) of backdoored DRL models in the clean environment without attackers. We can find that the model of P-EADQN has about 7\% performance degradation relative to the clean one of EADQN. The models of P-DRQN and P-DAGQN obtain the performance that is comparable to that of their clean models. Among four training cases in Table \ref{tab:aperformance}, the worst PVR of DRQN and DAGQN without and with backdoors is 4.4\% and 3.1\%, respectively. Hence, the backdoored DAGQN model is better than  P-DRQN and P-EADQN and is comparable to its clean model. In Table \ref{tab:aperformance}, by randomly injecting malicious driving behaviors following triggers, we compare the ASR and durations to crash (DC) of these backdoored models after these trigger behaviors are present. Our P-DAGQN can obtain 100\% ASR both in the highway and twoway scenarios. Since the CR achieved by DAGQN in the roundabout scenario is higher than other scenarios and these failing cases interfere with backdoor learning, P-DAGQN can only obtain 98.5\% ASR. In these four backdoor cases, P-EADQN almost fails to generate malicious actions after the trigger is present. The ASR achieved by P-DRQN is better than P-EADQN, but is still far from acceptable for efficient attack deployment. This phenomenon has already been shown in Fig. \ref{fig:conv} (a) that the rewards both of P-DAGQN and P-DRQN models are higher than that of clean models, but P-EADQN has no such curves. Additionally, P-DAGQN has lower DC than other algorithms, so that P-DAGQN can achieve better stealthiness in the term of poison rate and can crash the AV more quickly. That is the reason why P-DAGQN has higher average rewards and lower running durations than DAGQN as shown in Fig. \ref{fig:conv}.

In summary, DRQN can capture temporal driving features by RNN and EADQN can capture vehicle-to-ego features by the attention mechanism. Hence, DRQN and EADQN can achieve the performance better than DQN. Since our DAGQN can capture both spatial and temporal features, the AV controlled by DAGQN policies outperforms than others in these three driving scenarios. However, this excellent performance is a double-edged sword that it can be also reflected in the injection of backdoor attacks.

\begin{figure}[!t]
\centering
  \vspace{-4mm}
\begin{minipage}[c]{1\columnwidth}
    \centering
    \subfloat[Turn left (Prediction vs. GT)]{\includegraphics[width=0.45\columnwidth]{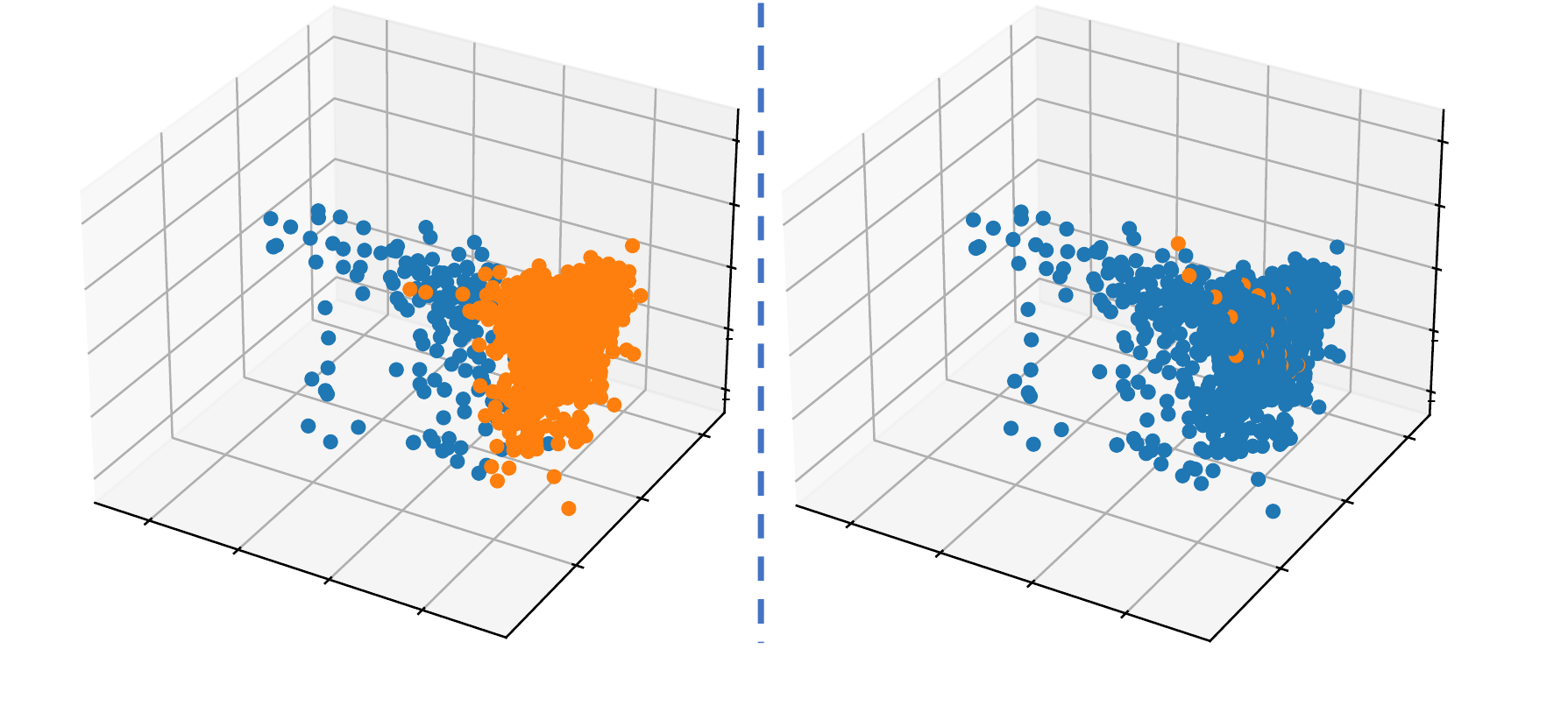}}
    \hspace{3mm}
    \subfloat[Turn right (Prediction vs. GT)]{\includegraphics[width=0.45\columnwidth]{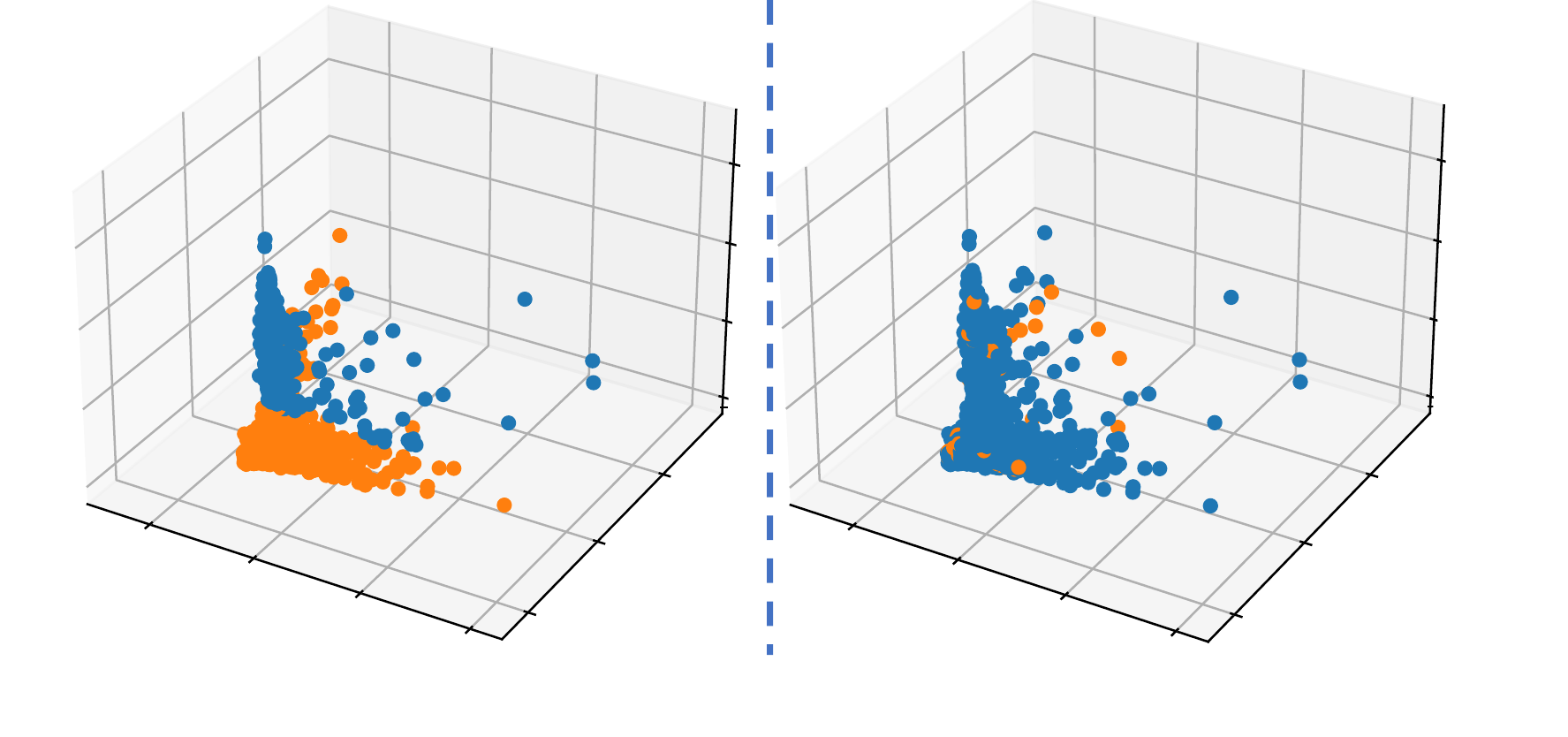}}
    \vspace{-3mm}
  \end{minipage}

  \begin{minipage}[c]{1\columnwidth}
      \centering
      \subfloat[Speed up (Prediction vs. GT)]{\includegraphics[width=0.45\columnwidth]{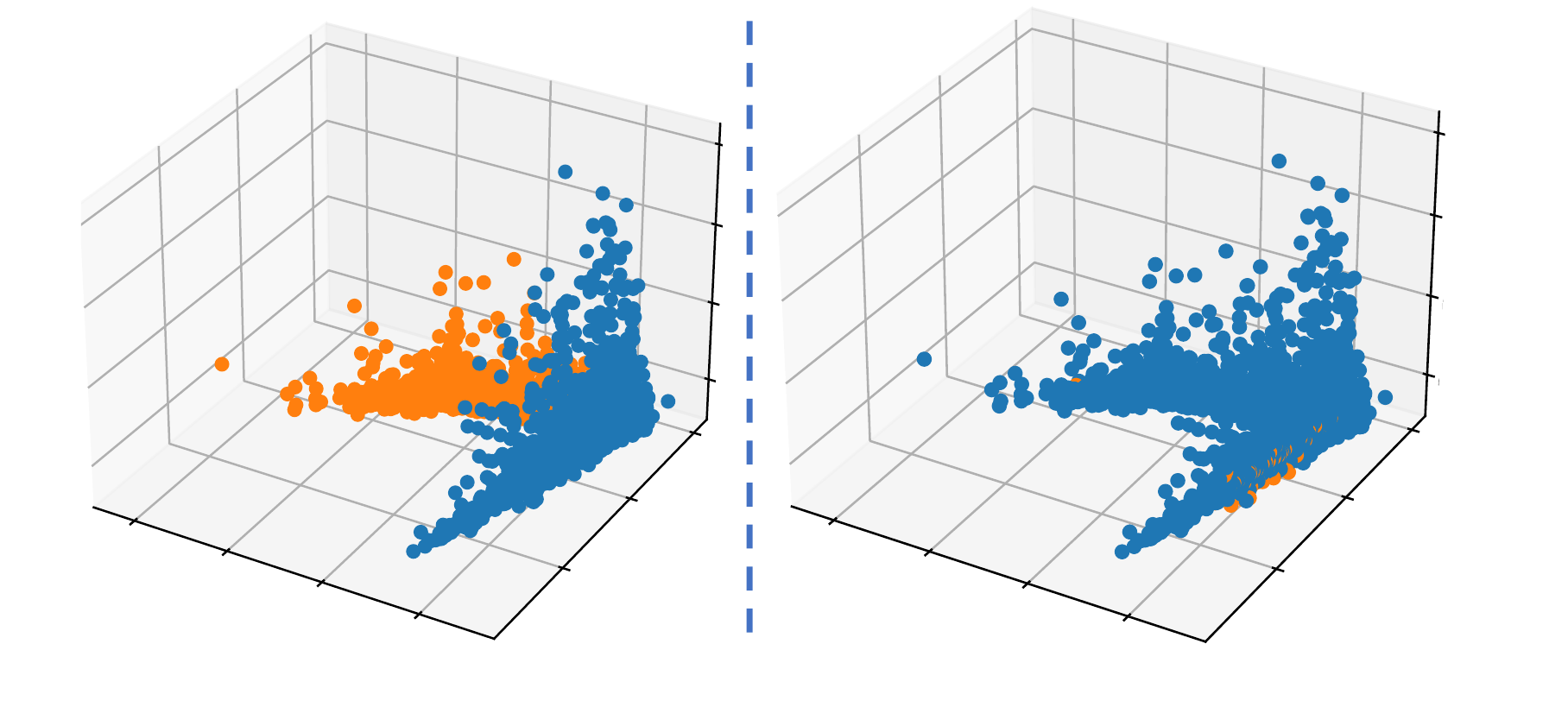}}
      \hspace{3mm}
      \subfloat[Slow down (Prediction vs. GT)]{\includegraphics[width=0.45\columnwidth]{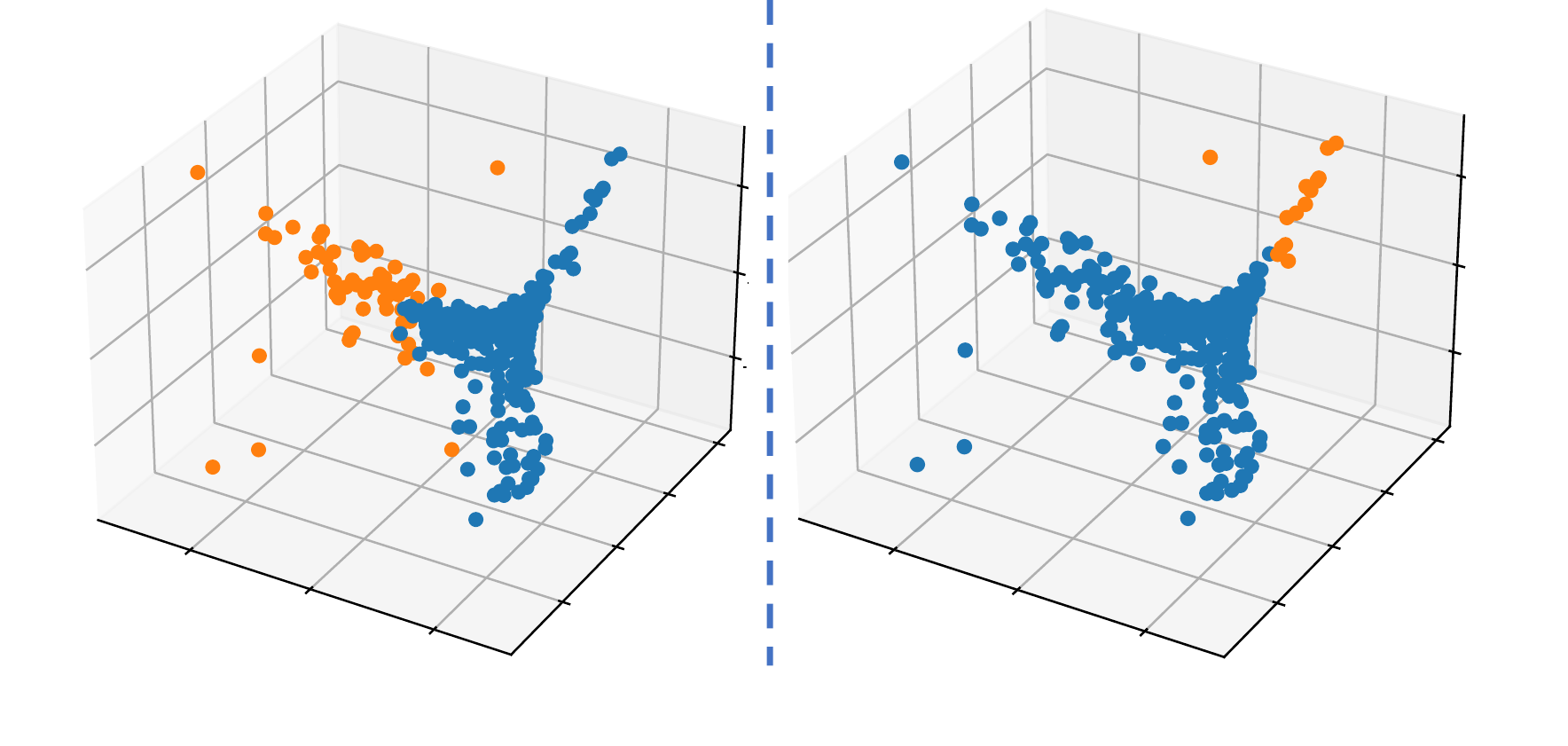}}
    \end{minipage}
\begin{minipage}[c]{1\columnwidth}
    \centering
    \subfloat[Cruise (Prediction vs. GT)]{\includegraphics[width=0.45\columnwidth]{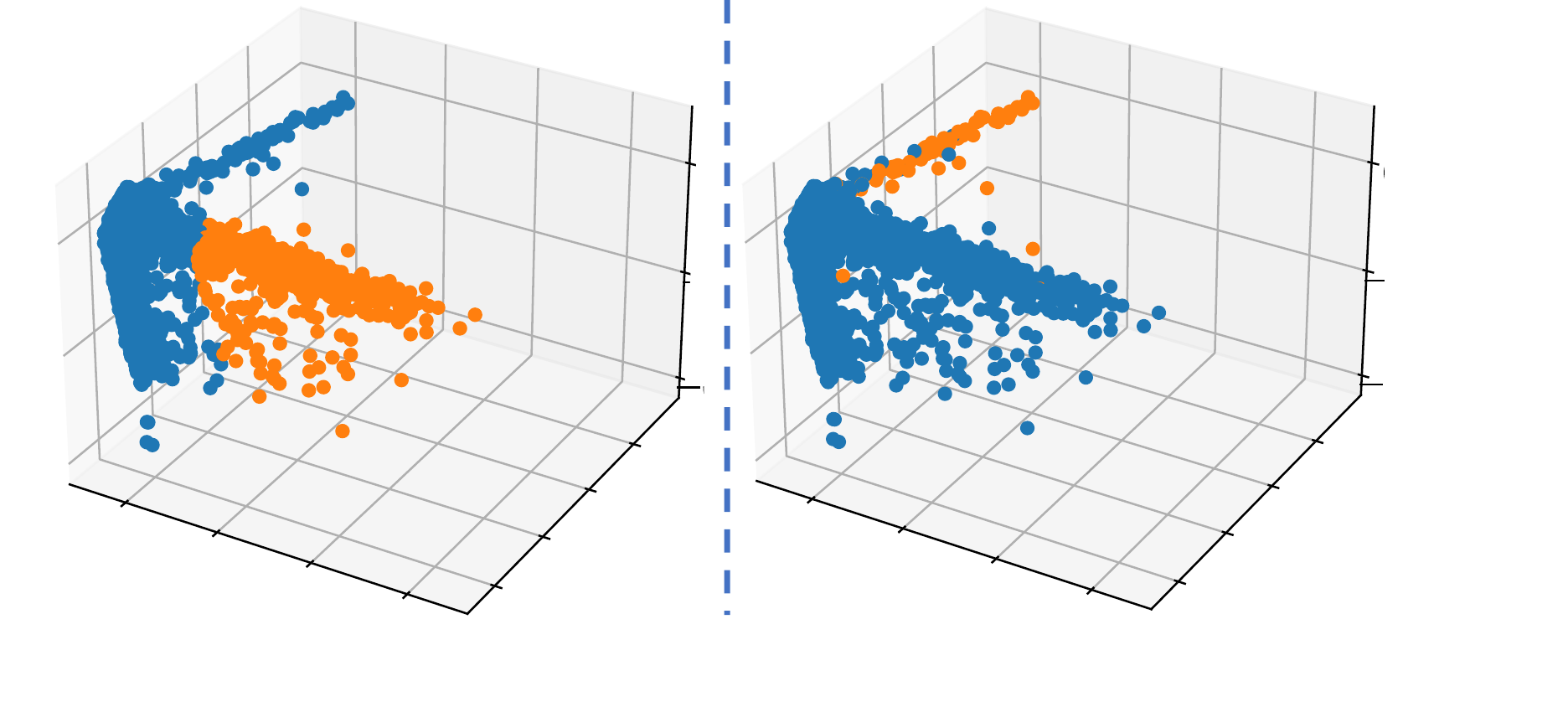}}
  \end{minipage}

  \caption{Activation clustering. ``Predication'' is generated by the activation clustering. ``GT'' means groundtruth.}
  \label{fig:ac}
\end{figure}

\vspace{-2mm}
\subsection{Defense}
We further evaluate whether existing SOTA defense approaches implemented in \cite{art} (spectral signatures \cite{tran2018spectral} and activation clustering \cite{chen2018detecting}) can identify a Trojan in our trained models. We run the highway-env simulator with our trained model to generate 24000 samples, in which there are 494 poisonous samples. We divide these samples into five sets according to their action types and present their spectral signature in Fig. \ref{fig:spec}. We can find that the distribution of poisonous samples are covered by the distribution of clean samples and thus, these samples are not distinguishable for backdoors. Besides, we extract activations of the penultimate layer of our trained model, perform independent component analysis to reduce data dimensionality, and cluster them using k-means with k = 2. We show the results activation clustering in Fig. \ref{fig:ac}, in which we can see that the activations from the poisonous samples are not distinguishable with the clean ones, and even the predicted results are completely opposite to the ground truth (e.g., Fig. \ref{fig:ac} (d) and (e)).

Besides a minuscule amount of states are poisoned, the fundamental reason behind these failing detection results is that our backdoor attack are much different than Trojans in supervised learning (e.g., image classification). A poisonous sample and a malicious action have the one-to-one relationship in supervised learning. But in our case, poisonous samples and malicious actions are distributed in a series of states. Specifically, our trigger is a combination of spatial and temporal features, rather than a single instant state, and our malicious actions are learned using specified reward functions, rather than fixed. A possible defense mechanism needs to have the capacity of capturing spatial and temporal features over sequential states, rather than individual state.

\section{Conclusions and Future}
\label{sec:conclusion}
In this paper, we study the security threats to DRL-augment AD systems using spatio-temporal traffic features. We first propose a spatio-temporal DNN model based on GRU and attention mechanism for training DQN-based policies for AD tasks. Then, we design a novel stealthy and practice backdoor attack on DRL-augment AD systems that hides its trigger into the spatial and temporal relevances of a sequence of states rather than a single state like existing DNN backdoor. Through numerous experiments, we show that while capturing spatio-temporal traffic features can improve the efficient and safety of the DRL-augment AD system, it suffer from backdoor threats for AD tasks. Our experiments show that our designed backdoor attacks on AD tasks have low performance degradation and high attack success rate, and are sustainable under SOTA backdoor defenses. Hence, in the future, we aim to design defense mechanisms suitable for DRL backdoors.

\bibliographystyle{IEEEtran}
\bibliography{IEEEabrv,ref}

\end{document}